\documentclass[twocolumn,english,prl,showpacs]{revtex4-1}
\usepackage[colorlinks=true,urlcolor=blue,citecolor=blue,linkcolor=blue]{hyperref}
\usepackage[T1]{fontenc}
\usepackage[latin9]{inputenc}
\usepackage{amssymb}
\usepackage{graphicx}
\usepackage{amsmath,color}
\usepackage{mathrsfs}
\usepackage{float}
\usepackage{braket}
\usepackage{indentfirst}

\makeatletter

\def\journal #1, #2, #3, 1#4#5#6{{\sl #1~}{\bf #2}, #3 (1#4#5#6) }

\def\eqa{\begin{eqnarray}}
\def\eea{\end{eqnarray}}
\newcommand{\eq}{\begin{equation}}
\newcommand{\ee}{\end{equation}}

\newcommand{\Eq}[1]{Eq.~(\ref{#1})}

\newcommand{\Tr}{{\rm Tr}}

\makeatother

\usepackage{babel}

\begin{document}

\title{Fate of the Kondo Effect and Impurity Quantum Phase Transitions Through \\ the Lens of Fidelity Susceptibility}

\author{Lei Wang$^{1}$, Hiroshi Shinaoka$^{1,2}$ and Matthias Troyer$^{1}$}
\affiliation{$^{1}$Theoretische Physik, ETH Zurich, 8093 Zurich, Switzerland}
\affiliation{$^{2}$Department of Physics, University of Fribourg, 1700 Fribourg, Switzerland}

\begin{abstract}
The Kondo effect is an ubiquitous phenomenon appearing at low temperature in quantum confined systems coupled to a continuous bath. Efforts in understanding and controlling it have triggered important developments  across several disciplines of condensed matter physics. A recurring pattern in these studies is that the suppression of the Kondo effect often results in intriguing physical phenomena such as impurity quantum phase transitions or non-Fermi-liquid behavior. We show that the fidelity susceptibility is a sensitive indicator for such phenomena because it quantifies the sensitivity of the system's state with respect to its coupling to the bath. We demonstrate the power of fidelity susceptibility approach by using it to identify the crossover and quantum phase transitions in the one and two impurity Anderson models. 
\end{abstract}


\maketitle

The Kondo effect~\cite{hewson1997kondo} was first observed in 1934~\cite{1934Phy.....1.1115D} as a low temperature resistance minimum in gold and was explained by Kondo in 1964 by taking into account the scattering of conduction electrons and magnetic impurities~\cite{1964PThPh..32...37K}. However, Kondo's perturbative calculation exhibits an unphysical divergence of the resistance at zero temperature. Resolving the Kondo problem has ultimately led to significant theoretical progresses, including the formulation of the scaling laws~\cite{1970JPhC....3.2436A}, the development of the numerical renormalization group (NRG) method~\cite{Wilson:1975ve} and the application of phenomenological Fermi-liquid theory~\cite{nozieres1974fermi}, Bethe ansatz~\cite{1980JETPL..31..364V,1980PhRvL..45..379A} and boundary conformal field theory~\cite{Affleck95} to the quantum impurity problems. Experimental interests has increased in the late 1990s due to  breakthroughs in fabricating artificial nano-devices and creating  tunable Kondo effects~\cite{Kouwenhoven:2001wd, Glazman:1988ue, 1988PhRvL..61.1768N, Cronenwett:1998wc, GoldhaberGordon:1998vk, VanderWiel:2000uw}. Other relevant experimental systems include the dissipative two-state systems~\cite{1987RvMP...59....1L} and the heavy-fermion compounds~\cite{2001Natur.413..804S, 2008NatPh...4..186G}. 
Moreover, 
through the dynamical-mean-field-theory (DMFT) framework~\cite{Anonymous:z_AfEOwS, Maier:2005tj} a connection between quantum impurity problems and correlated lattice models has been established~\cite{Bulla:2006ek, Ferrero:2007bz}. 

A general description of the quantum impurity problems can be written as 
\begin{equation}
\hat{H}(\lambda) = \underbrace{\hat{H}_{\rm impurity} + \hat{H}_{\rm bath}}_{\hat{H}_{0}} + \lambda \hat{H}_{1},
\label{eq:split} 
\end{equation}
where $\hat{H}_{0}$ describes the quantum impurity together with a continuous bath, 
and the last term describes the coupling between them. We treat $\lambda$ as a parameter and aim to characterize the state of the quantum impurity as a function of  its coupling to the bath. 
The Kondo effect originates from the bath's tendency to screen the local moment formed on the quantum impurity. Renormalization group analysis shows that in the Kondo region the coupling strength flows to infinity at low energy~\cite{1970JPhC....3.2436A, Wilson:1975ve}, implying that the local moment will eventually get screened at low enough temperature even with an arbitrarily weak bare impurity-bath coupling strength. 

There are, however, various physical processes that can compete with the Kondo effect.
In the presence of such competitions the system may undergo an~\emph{impurity quantum phase transition} where a competing state (local moment, charge order etc.) takes over as the bath-impurity coupling $\lambda$ decreases. Suppression of the Kondo screening often leads to non-Fermi liquid behavior~\cite{bulla2003quantum, Vojta:2006gr}. 
However, different from the quantum phase transition in bulk systems~\cite{SachdevBook}, at such~\emph{impurity quantum critical point} only a \emph{non-extensive} term in the free energy becomes singular. It is not always straightforward to find local probes to identify the impurity phase transitions. 
The question arises how to diagnose and characterize such impurity quantum phase transitions in a general setting~\cite{bayat2014order}. 

In this Letter we argue that the fidelity susceptibility~\cite{You:2007ew, CamposVenuti:2007il} provides a general and direct probe for an impurity quantum phase transitions. With its origin in quantum information and in the differential geometry perspective of quantum states, the fidelity susceptibility does not depend on details of the physical systems. The quantum fidelity $F(\lambda_{1}, \lambda_{2})$ is defined as the overlap of two ground state wave-functions (or density matrices in the nonzero temperature case~\cite{Sirker:2010fu}) for coupling strengths $\lambda_{1}$ and $\lambda_{2}$. The fidelity susceptibility~\cite{You:2007ew, CamposVenuti:2007il} 


\begin{equation}
\chi_{F}(\lambda) = -\frac{\partial^{2} \ln F(\lambda, \lambda+\epsilon)}{\partial{\epsilon}^{2}} \bigg|_{\epsilon=0}
\label{eq:definition}
\end{equation}
typically exhibits a maximum at the phase boundary because the system's state changes drastically around the quantum critical point. 
Since the fidelity susceptibility also fulfills the scaling laws~\cite{CamposVenuti:2007il,Albuquerque:2010fv}, it is an effective tool to detect and characterize various quantum phase transitions, see Ref.~\cite{Gu:2010em} for a review.

Recently, some of us developed an efficient approach for calculating the fidelity susceptibility of quantum many-body systems~\cite{Wang:2015tw} using modern quantum Monte Carlo (QMC) methods~\cite{Sandvik:1991tn, PhysRevLett.77.5130,Prokofev:1998tc,Evertz:2003ch,Kawashima:2004clb, Rubtsov:2005iw, Werner:2006ko, Gull:2008cm, 1999PhRvL..82.4155R, PhysRevB.91.241118, PhysRevB.91.235151}.  Specializing this to  quantum impurity models one can perform an  
expansion for the bath-impurity coupling~\cite{Gull:2011jd} 
\begin{eqnarray}
Z 
& =& \sum_{k=0}^{\infty} \lambda^{k} \int_{0}^{\beta} d\tau_{1}\ldots\int_{\tau_{k-1}}^{\beta} d\tau_{k}  \times  \nonumber \\ & & \Tr\left[(-1)^{k} e^{-(\beta-\tau_{k})\hat{H}_{0}}\hat{H}_{1} \ldots  \hat{H}_{1}e^{ -\tau_{1}\hat{H}_{0} } \right] . 
\label{eq:ct-expansion}
\end{eqnarray}
Depending on details of the impurity Hamiltonian, various QMC algorithms can be used to sample \Eq{eq:ct-expansion}. For example, the continuous-time hybridization expansion algorithm (CT-HYB)~\cite{Werner:2006ko} solves (multi-orbital) Anderson impurity models, while the CT-J algorithm~\cite{Otsuki:2007ff} is suitable for the Kondo model and its multi-orbital generalization such as the Coqblin-Schrieffer model~\cite{Coqblin:1969wu}. 

Equation~(\ref{eq:ct-expansion}) also provides a conceptual framework to understand an impurity quantum phase transition through a quantum-classical mapping. The expansion can be formally interpreted as a grand canonical partition function of classical particles residing on a ring of length $\beta$. These particles represent the bath-impurity coupling events provided by the $\hat{H}_{1}$ terms and their number is controlled by the  coupling strength $\lambda$. 
Since \Eq{eq:ct-expansion} has the form of a fugacity expansion, an impurity quantum phase transition driven by $\lambda$ will manifest itself as a condensation phase transition of classical particles~\cite{PhysRev.87.404, PhysRev.87.410}. 

\begin{figure}[tbp]
\centering
\includegraphics[width=8cm]{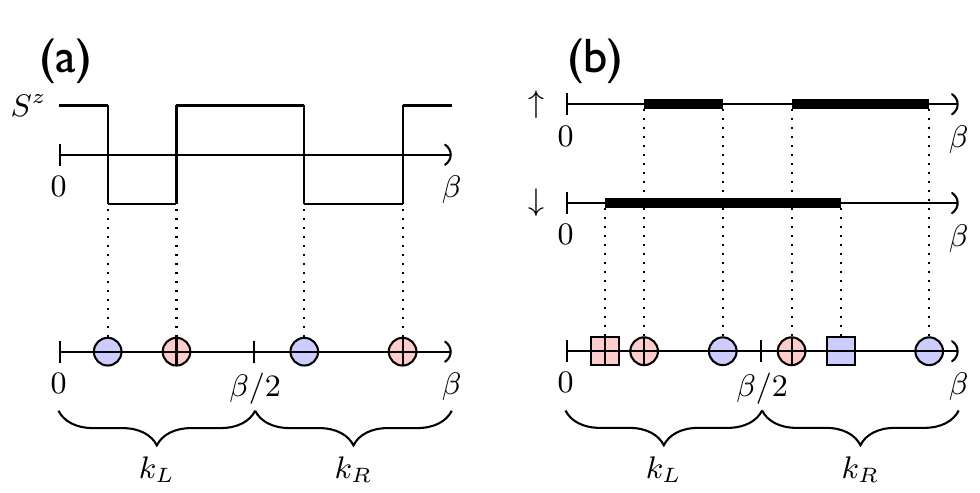}
\caption{(color online). (a) The Anderson-Yuval mapping of the Kondo model (\Eq{eq:kondo}) to a one-dimensional classical Coulomb gas. The magnetization of the quantum impurity flips in the imaginary time due to the coupling to the bath. The spin flips can be interpreted as positive (red circle) and negative (blue circle) charges distributed on a periodic ring. (b) 
The continuous-time hybridization-expansion QMC algorithm, in the same spirit, maps the single impurity Anderson model (\Eq{eq:siam}) to a classical statistical problem. 
The thick segments indicate the occupation of the spin up and down impurity levels. The endpoints of each segment represent the hybridization events where the electron hops in/out of the impurity,  which are treated as classical objects in the QMC sampling. In the both cases (a) and (b), the fidelity susceptibility is calculated as the covariance of $k_{L}$ and $k_{R}$, which count the number of bath-impurity coupling events in the two equal-bipartitions of the imaginary-time axis.   
}
\label{fig:concept}
\end{figure}

A concrete example of this general reasoning is provided by the Anderson-Yuval solution of the anisotropic Kondo model \cite{Anderson:1969wl, Anderson:1970th, schotte1970tomonaga, Anderson:1973hba} with Hamiltonian
\begin{eqnarray}
\hat{H}_\mathrm{Kondo} &=&  \sum_{\mathbf{k},\sigma} \epsilon_{\mathbf{k}}\hat{c}^{\dagger}_{\mathbf{k}\sigma}\hat{c}_{\mathbf{k}\sigma} + J_{z} \sum_{\mathbf{k}, \mathbf{k}^{\prime}}\hat{S}^{z} \left(\hat{c}^{\dagger}_{\mathbf{k}\uparrow}\hat{c}_{\mathbf{k}^{\prime}\uparrow}-\hat{c}^{\dagger}_{\mathbf{k}\downarrow}\hat{c}_{\mathbf{k}^{\prime}\downarrow} \right)  \nonumber \\  && + \lambda \sum_{\mathbf{k},\mathbf{k}^{\prime}} \left(\hat{S}^{+} \hat{c}_{\mathbf{k}\downarrow}^{\dagger} \hat{c}_{\mathbf{k}^{\prime}\uparrow} + h.c. \right) 
\label{eq:kondo}. 
\end{eqnarray}
The last term, which plays the role of $\lambda \hat{H}_{1}$ in \Eq{eq:split} describes the coupling of the local impurity spin to free electrons in the bath through spin-flips. An expansion in the form of \Eq{eq:ct-expansion} and integration out of the free fermions lead to Anderson and Yuval's mapping of the Kondo model to a one-dimensional classical Coulomb gas~\cite{Anderson:1969wl}.  As illustrated in Fig.~\ref{fig:concept}(a), the spin flips can be interpreted as alternating positive and negative charges interacting through a logarithmic Coulomb potential~\cite{Anderson:1969wl, Anderson:1970th, schotte1970tomonaga, Anderson:1973hba}. The Coulomb gas picture provides an intuitive understanding of the Kondo effect and the renormalization group flow~\cite{ Anderson:1970th}. For a ferromagnetic coupling $J_{z}<0$, the Coulomb gas exhibits a phase transition as the fugacity $\lambda$ changes. For small $\lambda<|J_{z}|$ these Coulomb charges are dilute and all associated in pairs, corresponding to the ferromagnetic Kondo state where the quantum impurity is spin polarized, while $\lambda>|J_{z}|$ corresponds to the antiferromagnetic Kondo state where the spin-flips are so frequent that the impurity shows no net magnetization, i.e. it is Kondo screened 
\footnote{Ref.~\cite{1971PhRvB...4.2228S} performed one of the first historic Carlo simulation of the Kondo model \Eq{eq:kondo} based on the Coulomb gas analogy. However the simulation was performed in a canonical ensemble with a fixed number of spin-flips, and thus has systematic errors in the high temperature region~\cite{Anderson:1973hba}.}. 

In the framework of \Eq{eq:ct-expansion}, the fidelity susceptibility (\ref{eq:definition}) can be readily calculated using a covariance estimator~\cite{Wang:2015tw}
\begin{equation}
\chi_{F} = \frac{\braket{k_{L}k_{R}} -\braket{k_{L}} \braket{k_{R}} }{2\lambda^{2}},  
\label{eq:fs}
\end{equation}
where $k_{L}$ and $k_{R}$ are the numbers of $\hat{H}_{1}$ operators in the two bipartitions of the imaginary-time axis. In the case of the Kondo model~\Eq{eq:kondo}, they correspond to the number of charges on either side of the bipartition, shown in the bottom of Fig.~\ref{fig:concept}(a). It is clear from Anderson and Yuval's classical Coulomb gas picture that the fidelity susceptibility estimator~(\ref{eq:fs}) captures the critical fluctuation upon a condensation phase transition therefore is able to signify the impurity phase transitions of the anisotropic Kondo model. 

For general quantum impurity models the estimator (\ref{eq:fs}) always quantifies the sensibility of the system's state with respect to the bath-impurity coupling and can  therefore be used to diagnose impurity quantum phase transitions. 
It is a generic probe of phase transition irrespective of physical details of the system. The singularity of the fidelity susceptibility upon a phase transition is also stronger than the second order derivative of the free energy (related to variance of the total expansion order)~\cite{Albuquerque:2010fv, Wang:2015tw}. Moreover, the fidelity susceptibility can also be used to inspect the crossover physics, even though there is no sharp phase boundary. 

As illustration we consider first the single impurity Anderson model~\cite{1961PhRv..124...41A}
\begin{eqnarray}
\hat{H}_\mathrm{SIAM} &=& \sum_{\mathbf{k},\sigma} \epsilon_{\mathbf{k}}\hat{c}^{\dagger}_{\mathbf{k}\sigma}\hat{c}_{\mathbf{k}\sigma} +\epsilon_{d}\sum_{\sigma}\hat{n}_{\sigma} +U \hat{n}_{\uparrow}\hat{n}_{\downarrow} \nonumber \\ &&+ \lambda \sum_{\mathbf{k},\sigma}\left(\hat{c}_{\mathbf{k}\sigma}^{\dagger} \hat{d}_{\sigma} + h.c.\right),
\label{eq:siam}
\end{eqnarray}
where $\hat{n}_{\sigma} =\hat{d}^{\dagger}_{\sigma} \hat{d}_{\sigma}$ is the impurity occupation number and the second line describes the hybridization of the impurity and the noninteracting bath with strength $\lambda$. We consider a noninteracting bath with semicircle density-of-states $\rho(\epsilon) = \sum_{\mathbf{k}} \delta(\epsilon-\epsilon_{\mathbf{k}})= \frac{2}{\pi D}\sqrt{1-(\epsilon/D)^{2}}$ with $D=2$ and choose $\epsilon_{d}=-U/2$ such that model is at the particle-hole symmetric point. As we tune the onsite interaction $U$ and the hybridization strength $\lambda$ there is a crossover from a local moment regime, where the spin of the singly occupied impurity is free to flip to the Kondo region, where the local moment is screened by the bath~\cite{KrishnaMurthy:1980wb}. 

We use the CT-HYB algorithm~\cite{Werner:2006ko} for our simulations and illustrate one Monte Carlo configuration in Fig.~\ref{fig:concept}(b). Each dashed line indicates a hybridization event, where the electron hops on or off   the  impurity site, thus changing the occupation (indicated by the thickness of the segments). The fidelity susceptibility \Eq{eq:fs} is easily measured by counting the number of hybridization events in a bipartition of the imaginary time axis, shown in the bottom of Fig.~\ref{fig:concept}(b).

\begin{figure}[t!]
\centering
\includegraphics[width=8cm]{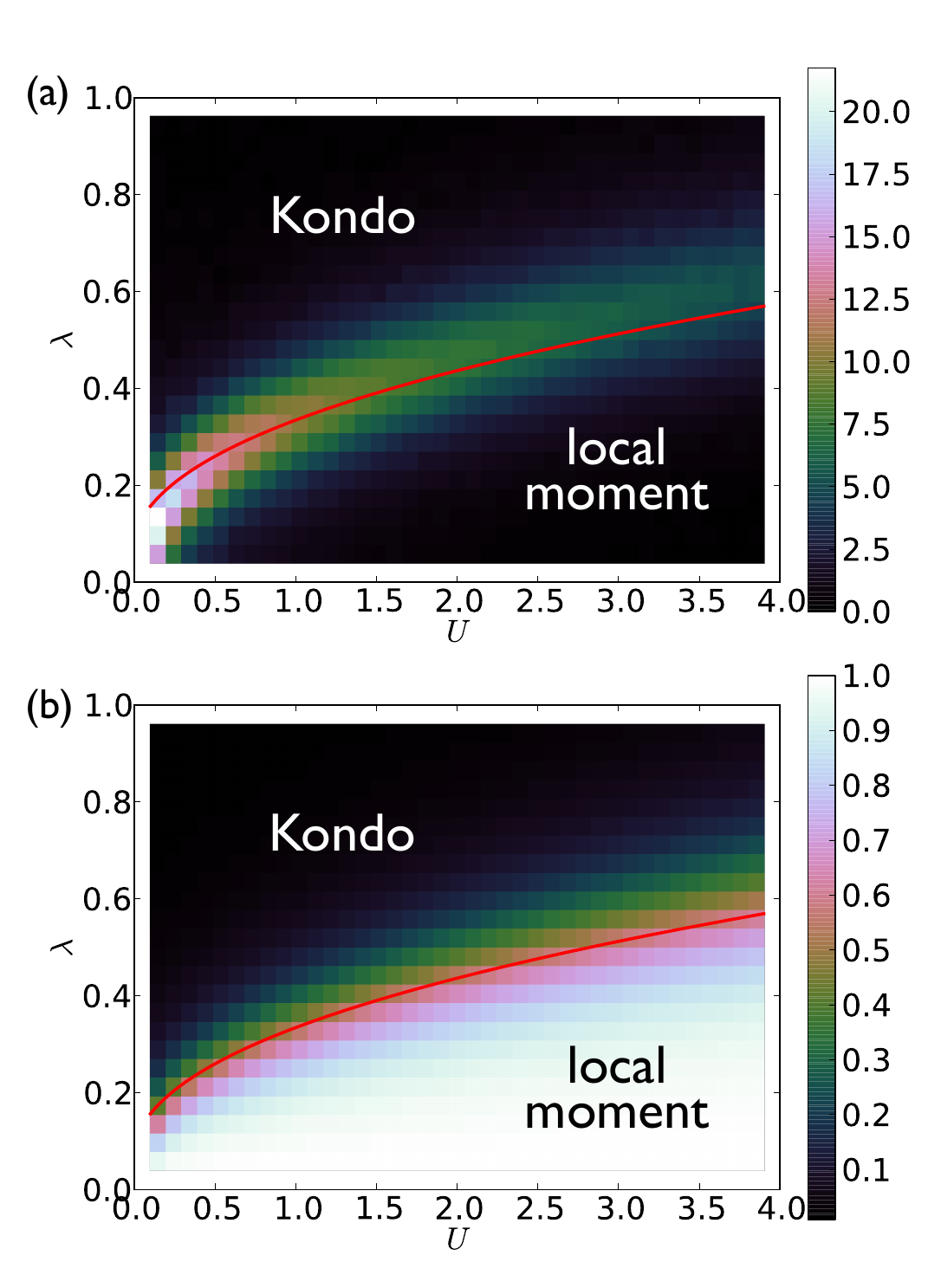}
\caption{(color online). Crossover from the local moment to the Kondo region in the single impurity Anderson model (\Eq{eq:siam}) revealed by (a) fidelity susceptibility (b) spin susceptibility $4T\chi_{s}$ defined in \Eq{eq:chis}. The red solid line shows the contour determined from the Kondo temperature $1/\beta = T_{K}(U,\lambda)$ \cite{1978PhRvL..40..416H,KrishnaMurthy:1980wb}.}
\label{fig:SIAM}
\end{figure}

Figure~\ref{fig:SIAM}(a) shows the fidelity susceptibility $\chi_{F}(\lambda)$ in the $U-\lambda$ plane with fixed inverse temperature $\beta=100$, where the peak indicates the crossover from local moment to Kondo region. 
This figure is a two-dimensional slice of the phase diagram of the Anderson impurity model sketched in the seminal NRG work of Ref.~\cite{KrishnaMurthy:1980wb} (Fig.~12). The red solid line shows the contour determined by the Kondo temperature $1/\beta=T_{K}(U, \lambda)= \lambda\sqrt{U}e^{-\pi U/(8\lambda^{2})}$~\cite{1978PhRvL..40..416H}. This  boundary agrees with the maxima of the fidelity susceptibility, showing that it indeed correctly captures the crossover physics. The peak of  fidelity susceptibility is higher at small $\lambda$ region, which is a manifestation of the Anderson orthogonality catastrophe~\cite{Anderson:1967tr}: even a weak coupling to the quantum impurity drastically change the state of the system. 

To further confirm the relevance of the peak of the fidelity susceptibility we calculate the local spin susceptibility 

\begin{equation}
\chi_{s}=\int_{0}^{\beta}d\tau  \braket{\hat{S}^{z}(\tau) \hat{S}^{z}(0)},
\label{eq:chis}
\end{equation}
where $\hat{S}^{z} = (\hat{n}_{\uparrow}-\hat{n}_{\downarrow})/2$ is the magnetization on the impurity. Figure~\ref{fig:SIAM}(b) shows that $4T\chi_{s}$, which corresponds to the effective moment on the impurity, changes from one in the local moment region to zero in the Kondo region. 
The crossover region agrees with the peak determined from the fidelity susceptibility in Fig.~\ref{fig:SIAM}(a). 


%

As a second example we consider the two-impurity Anderson model~\cite{PhysRevB.86.125134}
\begin{eqnarray}
\hat{H}_\mathrm{TIAM} &=& \sum_{\mathbf{k},\alpha,\sigma} \epsilon_{\mathbf{k}}\hat{c}^{\dagger}_{\mathbf{k}\alpha\sigma}\hat{c}_{\mathbf{k}\alpha\sigma} +\epsilon_{d}\sum_{\alpha,\sigma}\hat{n}_{\alpha\sigma}  +U\sum_{\alpha} \hat{n}_{\alpha\uparrow} \hat{n}_{\alpha\downarrow} \nonumber \\ &&  + J \hat{\pmb{S}}_{1}\cdot\hat{\pmb{S}}_{2}  + \lambda \sum_{\mathbf{k},\alpha,\sigma}\left(\hat{c}_{\mathbf{k}\alpha\sigma}^{\dagger} \hat{d}_{\alpha\sigma} + h.c.\right)   
\label{eq:tiam}, 
\end{eqnarray}
where $\alpha=\{1,2\}$ labels  two impurity sites with occupation number $\hat{n}_{\alpha \sigma}=\hat{d}^{\dagger}_{\alpha\sigma}\hat{d}_{\alpha \sigma}$. 
The impurities have the same local interaction $U$ and onsite energy $\epsilon_{d}=-U/2$. Each impurity is coupled to its own bath with the hybridization strength $\lambda$. The last term represents the Ruderman-Kittel-Kasuya-Yosida (RKKY)~\cite{PhysRev.96.99,kasuya1956theory,PhysRev.106.893}  interaction between magnetic impurities in a metal. In the absence of this term, each impurity is Kondo screened by its own bath for the choice of $\lambda = 1$ and $\beta=100$. However, the antiferromagnetic RKKY coupling $J>0$ favors a singlet formed between the two impurity spins, which competes with the Kondo screening and causes an impurity quantum phase transition~\cite{PhysRevLett.58.843, PhysRevLett.61.125}. Detailed studies of the two impurity Anderson (and Kondo) model have provided insights into various aspects of the Kondo~\cite{PhysRevB.52.9528,PhysRevLett.108.086405} and heavy fermion physics~\cite{PhysRevLett.95.016402}.

\begin{figure}[t]
\centering
\includegraphics[width=8cm]{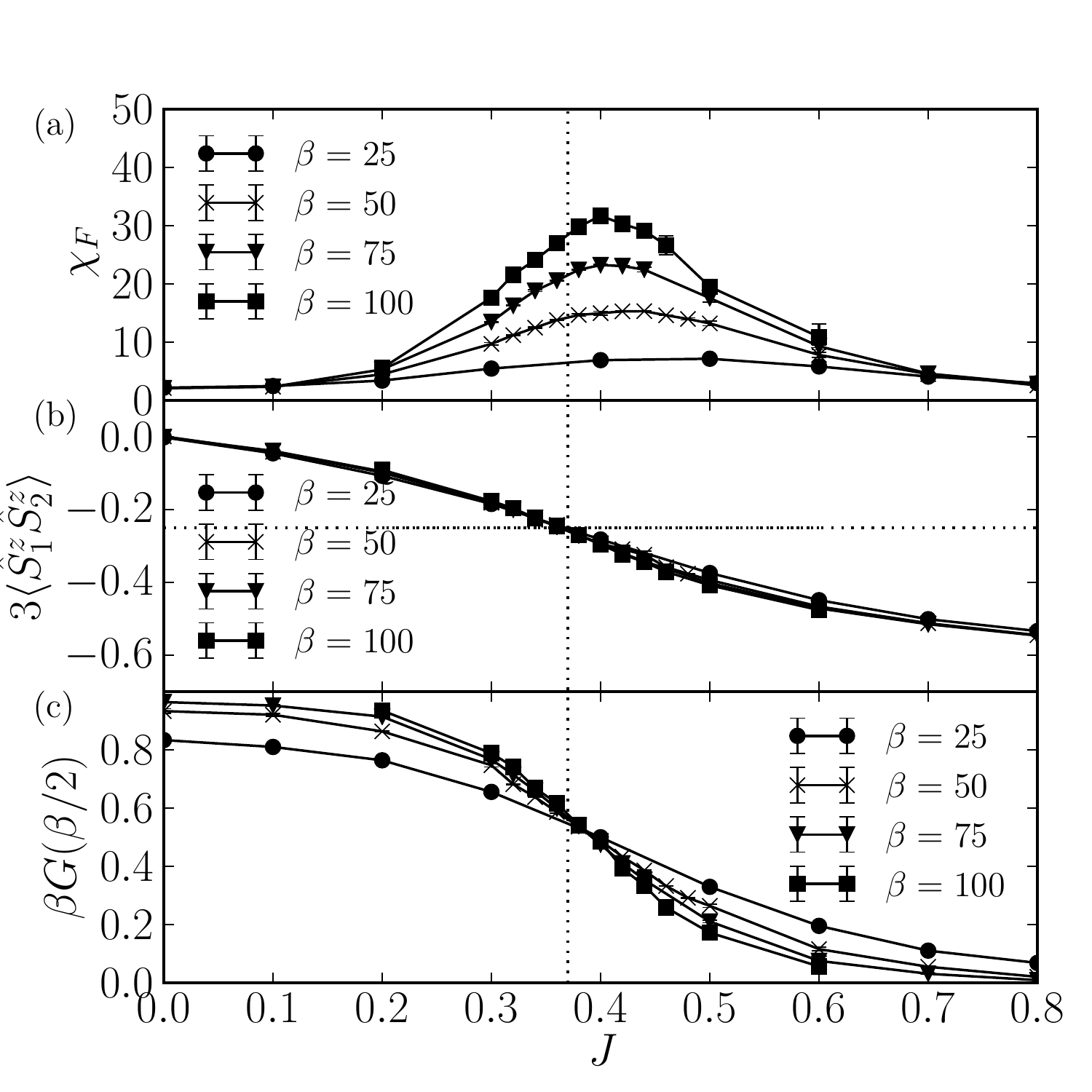}
\caption{(a) The fidelity susceptibility (b) equal-time spin-spin correlation (c) density-of-states at the Fermi level of the two-impurity Anderson model \Eq{eq:tiam} as a function of the inter-impurity RKKY coupling strength. The dashed vertical line indicates the critical point $J_c=0.37$ where the equal-time spin-spin correlation $ 3\braket{\hat{S}_{1}^{z}\hat{S}_{2}^{z}}=-0.25$~\cite{PhysRevLett.61.125,PhysRevB.40.324}.}
\label{fig:dimer}
\end{figure}


The simulation of this model goes beyond the segment picture illustrated in Fig.~\ref{fig:concept}(b), and we thus adopt an algorithm~\cite{Werner:2006iza, Gull:2011jd} suitable for {general interactions}. The fidelity susceptibility is still calculated in the same way, by simply counting the number of hybridization events. As  shown in Fig.~\ref{fig:dimer}(a), it exhibits an increasingly sharp peak as the inverse temperature $\beta$ increases. The peak location shifts towards the vertical dashed line, where the equal-time spin-spin correlation $\braket{\hat{\pmb{S}}_{1}\cdot\hat{\pmb{S}}_{2}}= 3\braket{\hat{S}_{1}^{z}\hat{S}_{2}^{z}}=-0.25$ in Fig.~\ref{fig:dimer}(b). According to  previous NRG studies~\cite{PhysRevLett.61.125,PhysRevB.40.324} the quantum critical point is right at the dashed line. Obviously, the fidelity susceptibility offers a better indication of the phase transition compared to the spin-spin correlations because the later quantity is featureless at the critical point and has much weaker temperature dependence. Figure~\ref{fig:dimer}(c) shows the density of states at the Fermi level, which decreases as the spin singlet state takes over the Kondo state in the large $J$ limit~\footnote{If there is a tunneling between the two impurities, the charge transfer will smear out the quantum phase transition into a crossover~\cite{PhysRevLett.97.166802}. Even in this case we still identify a peak in the fidelity susceptibility (not shown) that indicates the change of the system's state in the crossover region.}. 

\begin{figure}[t]
\centering
\includegraphics[width=9cm]{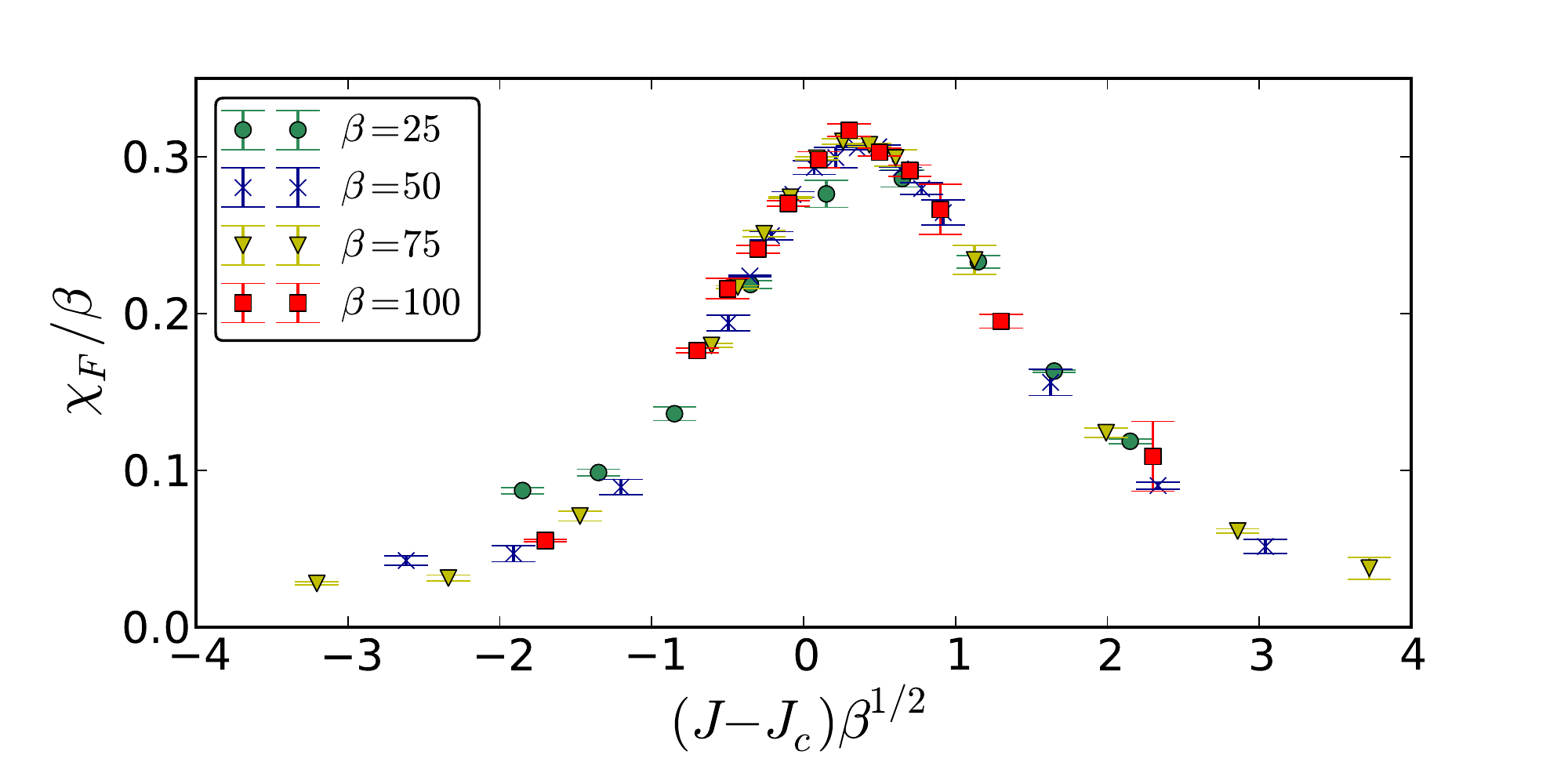}
\caption{(color online). Data collapse of the scaled fidelity susceptibility with $J_{c}=0.37$. The data are the same as the one in Fig.~\ref{fig:dimer}(a).}
\label{fig:scaling}
\end{figure}

We next perform a scaling analysis of the fidelity susceptibility close to the quantum critical point~\cite{Albuquerque:2010fv}. Since an infinite bath was assumed, the only finite dimension is the inverse temperature. Figure~\ref{fig:scaling} shows the scaled fidelity susceptibility $
\chi_{F}/\beta$ versus $(J-J_{c})\beta^{1/2}$ with $J_{c}=0.37$, which results in a good data collapse. Although the scaling form is chosen empirically according to the one for lattice systems~\cite{Albuquerque:2010fv}, the observed scaling exponents agree with the considerations of Ref.~\onlinecite{PhysRevLett.97.166802}. The observed data collapse suggests that the fidelity susceptibility not only captures the impurity quantum critical point but also the values of the critical exponents, which are an indicator for the  universality class of a quantum phase transitions. 

 
Our paper shows that the fidelity susceptibility is a versatile tool to probe and inspect phase transition and crossover physics in  quantum impurity models. It is readily acccessible in QMC simulations and it serves as a general purpose indicator for the breakdown of the Kondo effect. Conceptually our work exploits the intrinsic quantum to classical mapping of the quantum impurity models in the context of modern QMC approaches. 

Recent experimental and theoretical studies explore an even richer variety of complex quantum impurities and phase transitions, such as the interplay of Kondo effect and inter-impurity couplings~\cite{2001Sci...293.2221J, Liu:2010fz, Bork:2011bnc}, coupling to superconducting or Dirac fermion baths~\cite{Maurand:2012cb, Mebrahtu:2012hg, Boris:2011cu, Chen:2011jm, PhysRevB.88.045101} and the effect of multi-levels or multi-channels~\cite{Potok:2007bfa, Roch:2008is, Wang:2010dv, Goldstein:2010bo}, see Refs.~\cite{Vojta:2006gr, Rau:2013vf} for a review. The fidelity susceptibility will provide a valuable tool to discover rich physical phenomena in such settings. In a broader context, since the Kondo effect in the quantum impurity models is often linked to the Fermi liquid behavior in the framework of DMFT, calculating the fidelity susceptibility~\Eq{eq:fs} of the auxiliary quantum impurity problems may shed light on phase transitions of  correlated materials. 

We thank Ralf Bulla, Ninghua Tong and U.-J. Wiese for helpful discussions. Simulations were performed on the M\"{o}nch  and  Brutus clusters at ETH Zurich. We have used the ALPS  \verb|hybridization| application \cite{Hafermann:2013ilc} and ALPS libraries~\cite{BBauer:2011tz} for our simulations and data analysis. This work was supported by ERC Advanced Grant SIMCOFE, by the Swiss National Science Foundation through grant 200021E-149122 and the National Centers of Competence in Research QSIT and MARVEL, and by the DFG via FOR 1346 and the SNF Grant.  MT acknowledges hospitality of the Aspen Center for Physics, supported by NSF grant \# PHY-1066293. 

\bibliographystyle{apsrev4-1}
\bibliography{kondo_fs}

\begin{thebibliography}{86}%
\makeatletter
\providecommand \@ifxundefined [1]{%
 \@ifx{#1\undefined}
}%
\providecommand \@ifnum [1]{%
 \ifnum #1\expandafter \@firstoftwo
 \else \expandafter \@secondoftwo
 \fi
}%
\providecommand \@ifx [1]{%
 \ifx #1\expandafter \@firstoftwo
 \else \expandafter \@secondoftwo
 \fi
}%
\providecommand \natexlab [1]{#1}%
\providecommand \enquote  [1]{``#1''}%
\providecommand \bibnamefont  [1]{#1}%
\providecommand \bibfnamefont [1]{#1}%
\providecommand \citenamefont [1]{#1}%
\providecommand \href@noop [0]{\@secondoftwo}%
\providecommand \href [0]{\begingroup \@sanitize@url \@href}%
\providecommand \@href[1]{\@@startlink{#1}\@@href}%
\providecommand \@@href[1]{\endgroup#1\@@endlink}%
\providecommand \@sanitize@url [0]{\catcode `\\12\catcode `\$12\catcode
  `\&12\catcode `\#12\catcode `\^12\catcode `\_12\catcode `\%12\relax}%
\providecommand \@@startlink[1]{}%
\providecommand \@@endlink[0]{}%
\providecommand \url  [0]{\begingroup\@sanitize@url \@url }%
\providecommand \@url [1]{\endgroup\@href {#1}{\urlprefix }}%
\providecommand \urlprefix  [0]{URL }%
\providecommand \Eprint [0]{\href }%
\providecommand \doibase [0]{http://dx.doi.org/}%
\providecommand \selectlanguage [0]{\@gobble}%
\providecommand \bibinfo  [0]{\@secondoftwo}%
\providecommand \bibfield  [0]{\@secondoftwo}%
\providecommand \translation [1]{[#1]}%
\providecommand \BibitemOpen [0]{}%
\providecommand \bibitemStop [0]{}%
\providecommand \bibitemNoStop [0]{.\EOS\space}%
\providecommand \EOS [0]{\spacefactor3000\relax}%
\providecommand \BibitemShut  [1]{\csname bibitem#1\endcsname}%
\let\auto@bib@innerbib\@empty
\bibitem [{\citenamefont {Hewson}(1997)}]{hewson1997kondo}%
  \BibitemOpen
  \bibfield  {author} {\bibinfo {author} {\bibfnamefont {A.~C.}\ \bibnamefont
  {Hewson}},\ }\href@noop {} {\emph {\bibinfo {title} {The Kondo problem to
  heavy fermions}}}\ (\bibinfo  {publisher} {Cambridge university press},\
  \bibinfo {year} {1997})\BibitemShut {NoStop}%
\bibitem [{\citenamefont {de~Haas}\ \emph {et~al.}(1934)\citenamefont
  {de~Haas}, \citenamefont {de~Boer},\ and\ \citenamefont {van~d{\"e}n
  Berg}}]{1934Phy.....1.1115D}%
  \BibitemOpen
  \bibfield  {author} {\bibinfo {author} {\bibfnamefont {W.~J.}\ \bibnamefont
  {de~Haas}}, \bibinfo {author} {\bibfnamefont {J.}~\bibnamefont {de~Boer}}, \
  and\ \bibinfo {author} {\bibfnamefont {G.~J.}\ \bibnamefont {van~d{\"e}n
  Berg}},\ }\href
  {http://www.sciencedirect.com/science/article/pii/S0031891434803102}
  {\bibfield  {journal} {\bibinfo  {journal} {Physica}\ }\textbf {\bibinfo
  {volume} {1}},\ \bibinfo {pages} {1115} (\bibinfo {year} {1934})}\BibitemShut
  {NoStop}%
\bibitem [{\citenamefont {Kondo}(1964)}]{1964PThPh..32...37K}%
  \BibitemOpen
  \bibfield  {author} {\bibinfo {author} {\bibfnamefont {J.}~\bibnamefont
  {Kondo}},\ }\href
  {http://adsabs.harvard.edu/cgi-bin/nph-data_query?bibcode=1964PThPh..32...37K&link_type=ABSTRACT}
  {\bibfield  {journal} {\bibinfo  {journal} {Progress of Theoretical Physics}\
  }\textbf {\bibinfo {volume} {32}},\ \bibinfo {pages} {37} (\bibinfo {year}
  {1964})}\BibitemShut {NoStop}%
\bibitem [{\citenamefont {Anderson}(1970)}]{1970JPhC....3.2436A}%
  \BibitemOpen
  \bibfield  {author} {\bibinfo {author} {\bibfnamefont {P.~W.}\ \bibnamefont
  {Anderson}},\ }\href
  {http://adsabs.harvard.edu/cgi-bin/nph-data_query?bibcode=1970JPhC....3.2436A&link_type=EJOURNAL}
  {\bibfield  {journal} {\bibinfo  {journal} {J. Phys. C: Solid State Phys.}\
  }\textbf {\bibinfo {volume} {3}},\ \bibinfo {pages} {2436} (\bibinfo {year}
  {1970})}\BibitemShut {NoStop}%
\bibitem [{\citenamefont {Wilson}(1975)}]{Wilson:1975ve}%
  \BibitemOpen
  \bibfield  {author} {\bibinfo {author} {\bibfnamefont {K.~G.}\ \bibnamefont
  {Wilson}},\ }\href
  {http://journals.aps.org/rmp/abstract/10.1103/RevModPhys.47.773} {\bibfield
  {journal} {\bibinfo  {journal} {Rev. Mod. Phys.}\ }\textbf {\bibinfo {volume}
  {47}},\ \bibinfo {pages} {773} (\bibinfo {year} {1975})}\BibitemShut
  {NoStop}%
\bibitem [{\citenamefont {Nozieres}(1974)}]{nozieres1974fermi}%
  \BibitemOpen
  \bibfield  {author} {\bibinfo {author} {\bibfnamefont {P.}~\bibnamefont
  {Nozieres}},\ }\href@noop {} {\bibfield  {journal} {\bibinfo  {journal}
  {Journal of low temp{\'e}rature physics}\ }\textbf {\bibinfo {volume} {17}},\
  \bibinfo {pages} {31} (\bibinfo {year} {1974})}\BibitemShut {NoStop}%
\bibitem [{\citenamefont {Vigman}(1980)}]{1980JETPL..31..364V}%
  \BibitemOpen
  \bibfield  {author} {\bibinfo {author} {\bibfnamefont {P.~B.}\ \bibnamefont
  {Vigman}},\ }\href
  {http://adsabs.harvard.edu/cgi-bin/nph-data_query?bibcode=1980JETPL..31..364V&link_type=EJOURNAL}
  {\bibfield  {journal} {\bibinfo  {journal} {Journal of Experimental and
  Theoretical Physics Letters}\ }\textbf {\bibinfo {volume} {31}},\ \bibinfo
  {pages} {364} (\bibinfo {year} {1980})}\BibitemShut {NoStop}%
\bibitem [{\citenamefont {Andrei}(1980)}]{1980PhRvL..45..379A}%
  \BibitemOpen
  \bibfield  {author} {\bibinfo {author} {\bibfnamefont {N.}~\bibnamefont
  {Andrei}},\ }\href
  {http://adsabs.harvard.edu/cgi-bin/nph-data_query?bibcode=1980PhRvL..45..379A&link_type=ABSTRACT}
  {\bibfield  {journal} {\bibinfo  {journal} {Phys. Rev. Lett.}\ }\textbf
  {\bibinfo {volume} {45}},\ \bibinfo {pages} {379} (\bibinfo {year}
  {1980})}\BibitemShut {NoStop}%
\bibitem [{\citenamefont {Affleck}(1995)}]{Affleck95}%
  \BibitemOpen
  \bibfield  {author} {\bibinfo {author} {\bibfnamefont {I.}~\bibnamefont
  {Affleck}},\ }\href@noop {} {\bibfield  {journal} {\bibinfo  {journal} {arXiv
  preprint cond-mat/9512099}\ } (\bibinfo {year} {1995})}\BibitemShut {NoStop}%
\bibitem [{\citenamefont {Kouwenhoven}\ and\ \citenamefont
  {Glazman}(2001)}]{Kouwenhoven:2001wd}%
  \BibitemOpen
  \bibfield  {author} {\bibinfo {author} {\bibfnamefont {L.}~\bibnamefont
  {Kouwenhoven}}\ and\ \bibinfo {author} {\bibfnamefont {L.}~\bibnamefont
  {Glazman}},\ }\href
  {http://iopscience.iop.org/pwa/full/pwa-pdf/14/1/phwv14i1a28.pdf} {\bibfield
  {journal} {\bibinfo  {journal} {Physics world}\ }\textbf {\bibinfo {volume}
  {14}},\ \bibinfo {pages} {33} (\bibinfo {year} {2001})}\BibitemShut {NoStop}%
\bibitem [{\citenamefont {Glazman}\ and\ \citenamefont
  {Raikh}(1988)}]{Glazman:1988ue}%
  \BibitemOpen
  \bibfield  {author} {\bibinfo {author} {\bibfnamefont {L.~I.}\ \bibnamefont
  {Glazman}}\ and\ \bibinfo {author} {\bibfnamefont {M.~E.}\ \bibnamefont
  {Raikh}},\ }\href {http://www.jetpletters.ac.ru/ps/1095/article_16538.pdf}
  {\bibfield  {journal} {\bibinfo  {journal} {Jetp Lett.}\ }\textbf {\bibinfo
  {volume} {47}},\ \bibinfo {pages} {452} (\bibinfo {year} {1988})}\BibitemShut
  {NoStop}%
\bibitem [{\citenamefont {Ng}\ and\ \citenamefont
  {Lee}(1988)}]{1988PhRvL..61.1768N}%
  \BibitemOpen
  \bibfield  {author} {\bibinfo {author} {\bibfnamefont {T.-K.}\ \bibnamefont
  {Ng}}\ and\ \bibinfo {author} {\bibfnamefont {P.~A.}\ \bibnamefont {Lee}},\
  }\href
  {http://adsabs.harvard.edu/cgi-bin/nph-data_query?bibcode=1988PhRvL..61.1768N&link_type=ABSTRACT}
  {\bibfield  {journal} {\bibinfo  {journal} {Phys. Rev. Lett.}\ }\textbf
  {\bibinfo {volume} {61}},\ \bibinfo {pages} {1768} (\bibinfo {year}
  {1988})}\BibitemShut {NoStop}%
\bibitem [{\citenamefont {Cronenwett}\ \emph {et~al.}(1998)\citenamefont
  {Cronenwett}, \citenamefont {Oosterkamp},\ and\ \citenamefont
  {Kouwenhoven}}]{Cronenwett:1998wc}%
  \BibitemOpen
  \bibfield  {author} {\bibinfo {author} {\bibfnamefont {S.~M.}\ \bibnamefont
  {Cronenwett}}, \bibinfo {author} {\bibfnamefont {T.~H.}\ \bibnamefont
  {Oosterkamp}}, \ and\ \bibinfo {author} {\bibfnamefont {L.~P.}\ \bibnamefont
  {Kouwenhoven}},\ }\href
  {http://www.sciencemag.org/content/281/5376/540.short} {\bibfield  {journal}
  {\bibinfo  {journal} {Science}\ }\textbf {\bibinfo {volume} {281}},\ \bibinfo
  {pages} {540} (\bibinfo {year} {1998})}\BibitemShut {NoStop}%
\bibitem [{\citenamefont {Goldhaber-Gordon}\ \emph {et~al.}(1998)\citenamefont
  {Goldhaber-Gordon}, \citenamefont {Shtrikman}, \citenamefont {Mahalu},
  \citenamefont {Abusch-Magder}, \citenamefont {Meirav},\ and\ \citenamefont
  {Kastner}}]{GoldhaberGordon:1998vk}%
  \BibitemOpen
  \bibfield  {author} {\bibinfo {author} {\bibfnamefont {D.}~\bibnamefont
  {Goldhaber-Gordon}}, \bibinfo {author} {\bibfnamefont {H.}~\bibnamefont
  {Shtrikman}}, \bibinfo {author} {\bibfnamefont {D.}~\bibnamefont {Mahalu}},
  \bibinfo {author} {\bibfnamefont {D.}~\bibnamefont {Abusch-Magder}}, \bibinfo
  {author} {\bibfnamefont {U.}~\bibnamefont {Meirav}}, \ and\ \bibinfo {author}
  {\bibfnamefont {M.~A.}\ \bibnamefont {Kastner}},\ }\href
  {http://www.nature.com/nature/journal/v391/n6663/abs/391156a0.html}
  {\bibfield  {journal} {\bibinfo  {journal} {Nature}\ }\textbf {\bibinfo
  {volume} {391}},\ \bibinfo {pages} {156} (\bibinfo {year}
  {1998})}\BibitemShut {NoStop}%
\bibitem [{\citenamefont {Van~der Wiel}\ \emph {et~al.}(2000)\citenamefont
  {Van~der Wiel}, \citenamefont {De~Franceschi}, \citenamefont {Fujisawa},
  \citenamefont {Elzerman}, \citenamefont {Tarucha},\ and\ \citenamefont
  {Kouwenhoven}}]{VanderWiel:2000uw}%
  \BibitemOpen
  \bibfield  {author} {\bibinfo {author} {\bibfnamefont {W.~G.}\ \bibnamefont
  {Van~der Wiel}}, \bibinfo {author} {\bibfnamefont {S.}~\bibnamefont
  {De~Franceschi}}, \bibinfo {author} {\bibfnamefont {T.}~\bibnamefont
  {Fujisawa}}, \bibinfo {author} {\bibfnamefont {J.~M.}\ \bibnamefont
  {Elzerman}}, \bibinfo {author} {\bibfnamefont {S.}~\bibnamefont {Tarucha}}, \
  and\ \bibinfo {author} {\bibfnamefont {L.~P.}\ \bibnamefont {Kouwenhoven}},\
  }\href {http://www.sciencemag.org/content/289/5487/2105.short} {\bibfield
  {journal} {\bibinfo  {journal} {Science}\ }\textbf {\bibinfo {volume}
  {289}},\ \bibinfo {pages} {2105} (\bibinfo {year} {2000})}\BibitemShut
  {NoStop}%
\bibitem [{\citenamefont {Leggett}\ \emph {et~al.}(1987)\citenamefont
  {Leggett}, \citenamefont {Chakravarty}, \citenamefont {Dorsey}, \citenamefont
  {Fisher}, \citenamefont {Garg},\ and\ \citenamefont
  {Zwerger}}]{1987RvMP...59....1L}%
  \BibitemOpen
  \bibfield  {author} {\bibinfo {author} {\bibfnamefont {A.~J.}\ \bibnamefont
  {Leggett}}, \bibinfo {author} {\bibfnamefont {S.}~\bibnamefont
  {Chakravarty}}, \bibinfo {author} {\bibfnamefont {A.~T.}\ \bibnamefont
  {Dorsey}}, \bibinfo {author} {\bibfnamefont {M.~P.~A.}\ \bibnamefont
  {Fisher}}, \bibinfo {author} {\bibfnamefont {A.}~\bibnamefont {Garg}}, \ and\
  \bibinfo {author} {\bibfnamefont {W.}~\bibnamefont {Zwerger}},\ }\href
  {http://adsabs.harvard.edu/cgi-bin/nph-data_query?bibcode=1987RvMP...59....1L&link_type=ABSTRACT}
  {\bibfield  {journal} {\bibinfo  {journal} {Rev. Mod. Phys.}\ }\textbf
  {\bibinfo {volume} {59}},\ \bibinfo {pages} {1} (\bibinfo {year}
  {1987})}\BibitemShut {NoStop}%
\bibitem [{\citenamefont {Si}\ \emph {et~al.}(2001)\citenamefont {Si},
  \citenamefont {Rabello}, \citenamefont {Ingersent},\ and\ \citenamefont
  {Smith}}]{2001Natur.413..804S}%
  \BibitemOpen
  \bibfield  {author} {\bibinfo {author} {\bibfnamefont {Q.}~\bibnamefont
  {Si}}, \bibinfo {author} {\bibfnamefont {S.}~\bibnamefont {Rabello}},
  \bibinfo {author} {\bibfnamefont {K.}~\bibnamefont {Ingersent}}, \ and\
  \bibinfo {author} {\bibfnamefont {J.~L.}\ \bibnamefont {Smith}},\ }\href
  {http://adsabs.harvard.edu/cgi-bin/nph-data_query?bibcode=2001Natur.413..804S&link_type=ABSTRACT}
  {\bibfield  {journal} {\bibinfo  {journal} {Nature}\ }\textbf {\bibinfo
  {volume} {413}},\ \bibinfo {pages} {804} (\bibinfo {year}
  {2001})}\BibitemShut {NoStop}%
\bibitem [{\citenamefont {Gegenwart}\ \emph {et~al.}(2008)\citenamefont
  {Gegenwart}, \citenamefont {Si},\ and\ \citenamefont
  {Steglich}}]{2008NatPh...4..186G}%
  \BibitemOpen
  \bibfield  {author} {\bibinfo {author} {\bibfnamefont {P.}~\bibnamefont
  {Gegenwart}}, \bibinfo {author} {\bibfnamefont {Q.}~\bibnamefont {Si}}, \
  and\ \bibinfo {author} {\bibfnamefont {F.}~\bibnamefont {Steglich}},\ }\href
  {http://adsabs.harvard.edu/cgi-bin/nph-data_query?bibcode=2008NatPh...4..186G&link_type=ABSTRACT}
  {\bibfield  {journal} {\bibinfo  {journal} {Nature Physics}\ }\textbf
  {\bibinfo {volume} {4}},\ \bibinfo {pages} {186} (\bibinfo {year}
  {2008})}\BibitemShut {NoStop}%
\bibitem [{\citenamefont {Georges}\ \emph {et~al.}(1996)\citenamefont
  {Georges}, \citenamefont {Kotliar}, \citenamefont {Krauth},\ and\
  \citenamefont {Rozenberg}}]{Anonymous:z_AfEOwS}%
  \BibitemOpen
  \bibfield  {author} {\bibinfo {author} {\bibfnamefont {A.}~\bibnamefont
  {Georges}}, \bibinfo {author} {\bibfnamefont {G.}~\bibnamefont {Kotliar}},
  \bibinfo {author} {\bibfnamefont {W.}~\bibnamefont {Krauth}}, \ and\ \bibinfo
  {author} {\bibfnamefont {M.~J.}\ \bibnamefont {Rozenberg}},\ }\href
  {http://rmp.aps.org/abstract/RMP/v68/i1/p13_1} {\bibfield  {journal}
  {\bibinfo  {journal} {Rev. Mod. Phys.}\ }\textbf {\bibinfo {volume} {68}},\
  \bibinfo {pages} {13} (\bibinfo {year} {1996})}\BibitemShut {NoStop}%
\bibitem [{\citenamefont {Maier}\ \emph {et~al.}(2005)\citenamefont {Maier},
  \citenamefont {Jarrell}, \citenamefont {Pruschke},\ and\ \citenamefont
  {Hettler}}]{Maier:2005tj}%
  \BibitemOpen
  \bibfield  {author} {\bibinfo {author} {\bibfnamefont {T.}~\bibnamefont
  {Maier}}, \bibinfo {author} {\bibfnamefont {M.}~\bibnamefont {Jarrell}},
  \bibinfo {author} {\bibfnamefont {T.}~\bibnamefont {Pruschke}}, \ and\
  \bibinfo {author} {\bibfnamefont {M.~H.}\ \bibnamefont {Hettler}},\ }\href
  {http://rmp.aps.org/abstract/RMP/v77/i3/p1027_1} {\bibfield  {journal}
  {\bibinfo  {journal} {Rev. Mod. Phys.}\ }\textbf {\bibinfo {volume} {77}},\
  \bibinfo {pages} {1027} (\bibinfo {year} {2005})}\BibitemShut {NoStop}%
\bibitem [{\citenamefont {Bulla}(2006)}]{Bulla:2006ek}%
  \BibitemOpen
  \bibfield  {author} {\bibinfo {author} {\bibfnamefont {R.}~\bibnamefont
  {Bulla}},\ }\href {\doibase 10.1080/14786430500070313} {\bibfield  {journal}
  {\bibinfo  {journal} {Philosophical Magazine}\ }\textbf {\bibinfo {volume}
  {86}},\ \bibinfo {pages} {1877} (\bibinfo {year} {2006})}\BibitemShut
  {NoStop}%
\bibitem [{\citenamefont {Ferrero}\ \emph {et~al.}(2007)\citenamefont
  {Ferrero}, \citenamefont {De~Leo}, \citenamefont {Lecheminant},\ and\
  \citenamefont {Fabrizio}}]{Ferrero:2007bz}%
  \BibitemOpen
  \bibfield  {author} {\bibinfo {author} {\bibfnamefont {M.}~\bibnamefont
  {Ferrero}}, \bibinfo {author} {\bibfnamefont {L.}~\bibnamefont {De~Leo}},
  \bibinfo {author} {\bibfnamefont {P.}~\bibnamefont {Lecheminant}}, \ and\
  \bibinfo {author} {\bibfnamefont {M.}~\bibnamefont {Fabrizio}},\ }\href
  {http://stacks.iop.org/0953-8984/19/i=43/a=433201?key=crossref.648ca942a071b7c5f8be0a3d67f822dc}
  {\bibfield  {journal} {\bibinfo  {journal} {J. Phys.: Condens. Matter}\
  }\textbf {\bibinfo {volume} {19}},\ \bibinfo {pages} {433201} (\bibinfo
  {year} {2007})}\BibitemShut {NoStop}%
\bibitem [{\citenamefont {Bulla}\ and\ \citenamefont
  {Vojta}(2003)}]{bulla2003quantum}%
  \BibitemOpen
  \bibfield  {author} {\bibinfo {author} {\bibfnamefont {R.}~\bibnamefont
  {Bulla}}\ and\ \bibinfo {author} {\bibfnamefont {M.}~\bibnamefont {Vojta}},\
  }in\ \href@noop {} {\emph {\bibinfo {booktitle} {Concepts in Electron
  Correlation}}}\ (\bibinfo  {publisher} {Springer},\ \bibinfo {year} {2003})\
  pp.\ \bibinfo {pages} {209--217}\BibitemShut {NoStop}%
\bibitem [{\citenamefont {Vojta}(2006)}]{Vojta:2006gr}%
  \BibitemOpen
  \bibfield  {author} {\bibinfo {author} {\bibfnamefont {M.}~\bibnamefont
  {Vojta}},\ }\href
  {http://www.tandfonline.com/doi/abs/10.1080/14786430500070396} {\bibfield
  {journal} {\bibinfo  {journal} {Philosophical Magazine}\ }\textbf {\bibinfo
  {volume} {86}},\ \bibinfo {pages} {1807} (\bibinfo {year}
  {2006})}\BibitemShut {NoStop}%
\bibitem [{\citenamefont {Sachdev}(2011)}]{SachdevBook}%
  \BibitemOpen
  \bibfield  {author} {\bibinfo {author} {\bibfnamefont {S.}~\bibnamefont
  {Sachdev}},\ }\href@noop {} {\emph {\bibinfo {title} {Quantum Phase
  Transition}}}\ (\bibinfo  {publisher} {Cambridge Univ. Press},\ \bibinfo
  {year} {2011})\BibitemShut {NoStop}%
\bibitem [{\citenamefont {Bayat}\ \emph {et~al.}(2014)\citenamefont {Bayat},
  \citenamefont {Johannesson}, \citenamefont {Bose},\ and\ \citenamefont
  {Sodano}}]{bayat2014order}%
  \BibitemOpen
  \bibfield  {author} {\bibinfo {author} {\bibfnamefont {A.}~\bibnamefont
  {Bayat}}, \bibinfo {author} {\bibfnamefont {H.}~\bibnamefont {Johannesson}},
  \bibinfo {author} {\bibfnamefont {S.}~\bibnamefont {Bose}}, \ and\ \bibinfo
  {author} {\bibfnamefont {P.}~\bibnamefont {Sodano}},\ }\href
  {http://dx.doi.org/10.1038/ncomms4784} {\bibfield  {journal} {\bibinfo
  {journal} {Nature communications}\ }\textbf {\bibinfo {volume} {5}} (\bibinfo
  {year} {2014})}\BibitemShut {NoStop}%
\bibitem [{\citenamefont {You}\ \emph {et~al.}(2007)\citenamefont {You},
  \citenamefont {Li},\ and\ \citenamefont {Gu}}]{You:2007ew}%
  \BibitemOpen
  \bibfield  {author} {\bibinfo {author} {\bibfnamefont {W.-L.}\ \bibnamefont
  {You}}, \bibinfo {author} {\bibfnamefont {Y.~W.}\ \bibnamefont {Li}}, \ and\
  \bibinfo {author} {\bibfnamefont {S.-J.}\ \bibnamefont {Gu}},\ }\href
  {http://link.aps.org/doi/10.1103/PhysRevE.76.022101} {\bibfield  {journal}
  {\bibinfo  {journal} {Phys. Rev. E}\ }\textbf {\bibinfo {volume} {76}},\
  \bibinfo {pages} {022101} (\bibinfo {year} {2007})}\BibitemShut {NoStop}%
\bibitem [{\citenamefont {Campos~Venuti}\ and\ \citenamefont
  {Zanardi}(2007)}]{CamposVenuti:2007il}%
  \BibitemOpen
  \bibfield  {author} {\bibinfo {author} {\bibfnamefont {L.}~\bibnamefont
  {Campos~Venuti}}\ and\ \bibinfo {author} {\bibfnamefont {P.}~\bibnamefont
  {Zanardi}},\ }\href {http://link.aps.org/doi/10.1103/PhysRevLett.99.095701}
  {\bibfield  {journal} {\bibinfo  {journal} {Phys. Rev. Lett.}\ }\textbf
  {\bibinfo {volume} {99}},\ \bibinfo {pages} {095701} (\bibinfo {year}
  {2007})}\BibitemShut {NoStop}%
\bibitem [{\citenamefont {Sirker}(2010)}]{Sirker:2010fu}%
  \BibitemOpen
  \bibfield  {author} {\bibinfo {author} {\bibfnamefont {J.}~\bibnamefont
  {Sirker}},\ }\href {http://link.aps.org/doi/10.1103/PhysRevLett.105.117203}
  {\bibfield  {journal} {\bibinfo  {journal} {Phys. Rev. Lett.}\ }\textbf
  {\bibinfo {volume} {105}},\ \bibinfo {pages} {117203} (\bibinfo {year}
  {2010})}\BibitemShut {NoStop}%
\bibitem [{\citenamefont {Albuquerque}\ \emph {et~al.}(2010)\citenamefont
  {Albuquerque}, \citenamefont {Alet}, \citenamefont {Sire},\ and\
  \citenamefont {Capponi}}]{Albuquerque:2010fv}%
  \BibitemOpen
  \bibfield  {author} {\bibinfo {author} {\bibfnamefont {A.~F.}\ \bibnamefont
  {Albuquerque}}, \bibinfo {author} {\bibfnamefont {F.}~\bibnamefont {Alet}},
  \bibinfo {author} {\bibfnamefont {C.}~\bibnamefont {Sire}}, \ and\ \bibinfo
  {author} {\bibfnamefont {S.}~\bibnamefont {Capponi}},\ }\href {\doibase
  10.1103/PhysRevB.81.064418} {\bibfield  {journal} {\bibinfo  {journal} {Phys.
  Rev. B}\ }\textbf {\bibinfo {volume} {81}},\ \bibinfo {pages} {064418}
  (\bibinfo {year} {2010})}\BibitemShut {NoStop}%
\bibitem [{\citenamefont {Gu}(2010)}]{Gu:2010em}%
  \BibitemOpen
  \bibfield  {author} {\bibinfo {author} {\bibfnamefont {S.-J.}\ \bibnamefont
  {Gu}},\ }\href
  {http://www.worldscientific.com/doi/abs/10.1142/S0217979210056335} {\bibfield
   {journal} {\bibinfo  {journal} {Int. J. Mod. Phys. B}\ }\textbf {\bibinfo
  {volume} {24}},\ \bibinfo {pages} {4371} (\bibinfo {year}
  {2010})}\BibitemShut {NoStop}%
\bibitem [{\citenamefont {Wang}\ \emph
  {et~al.}(2015{\natexlab{a}})\citenamefont {Wang}, \citenamefont {Liu},
  \citenamefont {Imri\v{s}ka}, \citenamefont {Ma},\ and\ \citenamefont
  {Troyer}}]{Wang:2015tw}%
  \BibitemOpen
  \bibfield  {author} {\bibinfo {author} {\bibfnamefont {L.}~\bibnamefont
  {Wang}}, \bibinfo {author} {\bibfnamefont {Y.-H.}\ \bibnamefont {Liu}},
  \bibinfo {author} {\bibfnamefont {J.}~\bibnamefont {Imri\v{s}ka}}, \bibinfo
  {author} {\bibfnamefont {P.~N.}\ \bibnamefont {Ma}}, \ and\ \bibinfo {author}
  {\bibfnamefont {M.}~\bibnamefont {Troyer}},\ }\href {\doibase
  10.1103/PhysRevX.5.031007} {\bibfield  {journal} {\bibinfo  {journal} {Phys.
  Rev. X}\ }\textbf {\bibinfo {volume} {5}},\ \bibinfo {pages} {031007}
  (\bibinfo {year} {2015}{\natexlab{a}})}\BibitemShut {NoStop}%
\bibitem [{\citenamefont {Sandvik}\ and\ \citenamefont
  {Kurkij{\"a}rvi}(1991)}]{Sandvik:1991tn}%
  \BibitemOpen
  \bibfield  {author} {\bibinfo {author} {\bibfnamefont {A.~W.}\ \bibnamefont
  {Sandvik}}\ and\ \bibinfo {author} {\bibfnamefont {J.}~\bibnamefont
  {Kurkij{\"a}rvi}},\ }\href
  {http://journals.aps.org/prb/abstract/10.1103/PhysRevB.43.5950} {\bibfield
  {journal} {\bibinfo  {journal} {Phys. Rev. B}\ }\textbf {\bibinfo {volume}
  {43}},\ \bibinfo {pages} {5950} (\bibinfo {year} {1991})}\BibitemShut
  {NoStop}%
\bibitem [{\citenamefont {Beard}\ and\ \citenamefont
  {Wiese}(1996)}]{PhysRevLett.77.5130}%
  \BibitemOpen
  \bibfield  {author} {\bibinfo {author} {\bibfnamefont {B.}~\bibnamefont
  {Beard}}\ and\ \bibinfo {author} {\bibfnamefont {U.-J.}\ \bibnamefont
  {Wiese}},\ }\href {\doibase 10.1103/PhysRevLett.77.5130} {\bibfield
  {journal} {\bibinfo  {journal} {Phys. Rev. Lett.}\ }\textbf {\bibinfo
  {volume} {77}},\ \bibinfo {pages} {5130} (\bibinfo {year}
  {1996})}\BibitemShut {NoStop}%
\bibitem [{\citenamefont {Prokof'ev}\ \emph {et~al.}(1998)\citenamefont
  {Prokof'ev}, \citenamefont {Svistunov},\ and\ \citenamefont
  {Tupitsyn}}]{Prokofev:1998tc}%
  \BibitemOpen
  \bibfield  {author} {\bibinfo {author} {\bibfnamefont {N.~V.}\ \bibnamefont
  {Prokof'ev}}, \bibinfo {author} {\bibfnamefont {B.~V.}\ \bibnamefont
  {Svistunov}}, \ and\ \bibinfo {author} {\bibfnamefont {I.~S.}\ \bibnamefont
  {Tupitsyn}},\ }\href {http://link.springer.com/article/10.1134/1.558661}
  {\bibfield  {journal} {\bibinfo  {journal} {Journal of Experimental and
  Theoretical Physics}\ }\textbf {\bibinfo {volume} {87}},\ \bibinfo {pages}
  {310} (\bibinfo {year} {1998})}\BibitemShut {NoStop}%
\bibitem [{\citenamefont {Evertz}(2003)}]{Evertz:2003ch}%
  \BibitemOpen
  \bibfield  {author} {\bibinfo {author} {\bibfnamefont {H.~G.}\ \bibnamefont
  {Evertz}},\ }\href {\doibase 10.1080/0001873021000049195} {\bibfield
  {journal} {\bibinfo  {journal} {Advances in Physics}\ }\textbf {\bibinfo
  {volume} {52}},\ \bibinfo {pages} {1} (\bibinfo {year} {2003})}\BibitemShut
  {NoStop}%
\bibitem [{\citenamefont {Kawashima}\ and\ \citenamefont
  {Harada}(2004)}]{Kawashima:2004clb}%
  \BibitemOpen
  \bibfield  {author} {\bibinfo {author} {\bibfnamefont {N.}~\bibnamefont
  {Kawashima}}\ and\ \bibinfo {author} {\bibfnamefont {K.}~\bibnamefont
  {Harada}},\ }\href {\doibase 10.1143/JPSJ.73.1379} {\bibfield  {journal}
  {\bibinfo  {journal} {J. Phys. Soc. Jpn.}\ }\textbf {\bibinfo {volume}
  {73}},\ \bibinfo {pages} {1379} (\bibinfo {year} {2004})}\BibitemShut
  {NoStop}%
\bibitem [{\citenamefont {Rubtsov}\ \emph {et~al.}(2005)\citenamefont
  {Rubtsov}, \citenamefont {Savkin},\ and\ \citenamefont
  {Lichtenstein}}]{Rubtsov:2005iw}%
  \BibitemOpen
  \bibfield  {author} {\bibinfo {author} {\bibfnamefont {A.}~\bibnamefont
  {Rubtsov}}, \bibinfo {author} {\bibfnamefont {V.}~\bibnamefont {Savkin}}, \
  and\ \bibinfo {author} {\bibfnamefont {A.}~\bibnamefont {Lichtenstein}},\
  }\href {\doibase 10.1103/PhysRevB.72.035122} {\bibfield  {journal} {\bibinfo
  {journal} {Phys. Rev. B}\ }\textbf {\bibinfo {volume} {72}},\ \bibinfo
  {pages} {035122} (\bibinfo {year} {2005})}\BibitemShut {NoStop}%
\bibitem [{\citenamefont {Werner}\ \emph {et~al.}(2006)\citenamefont {Werner},
  \citenamefont {Comanac}, \citenamefont {de' Medici}, \citenamefont {Troyer},\
  and\ \citenamefont {Millis}}]{Werner:2006ko}%
  \BibitemOpen
  \bibfield  {author} {\bibinfo {author} {\bibfnamefont {P.}~\bibnamefont
  {Werner}}, \bibinfo {author} {\bibfnamefont {A.}~\bibnamefont {Comanac}},
  \bibinfo {author} {\bibfnamefont {L.}~\bibnamefont {de' Medici}}, \bibinfo
  {author} {\bibfnamefont {M.}~\bibnamefont {Troyer}}, \ and\ \bibinfo {author}
  {\bibfnamefont {A.}~\bibnamefont {Millis}},\ }\href
  {http://link.aps.org/doi/10.1103/PhysRevLett.97.076405} {\bibfield  {journal}
  {\bibinfo  {journal} {Phys. Rev. Lett.}\ }\textbf {\bibinfo {volume} {97}},\
  \bibinfo {pages} {076405} (\bibinfo {year} {2006})}\BibitemShut {NoStop}%
\bibitem [{\citenamefont {Gull}\ \emph {et~al.}(2008)\citenamefont {Gull},
  \citenamefont {Werner}, \citenamefont {Parcollet},\ and\ \citenamefont
  {Troyer}}]{Gull:2008cm}%
  \BibitemOpen
  \bibfield  {author} {\bibinfo {author} {\bibfnamefont {E.}~\bibnamefont
  {Gull}}, \bibinfo {author} {\bibfnamefont {P.}~\bibnamefont {Werner}},
  \bibinfo {author} {\bibfnamefont {O.}~\bibnamefont {Parcollet}}, \ and\
  \bibinfo {author} {\bibfnamefont {M.}~\bibnamefont {Troyer}},\ }\href
  {\doibase 10.1209/0295-5075/82/57003} {\bibfield  {journal} {\bibinfo
  {journal} {EPL}\ }\textbf {\bibinfo {volume} {82}},\ \bibinfo {pages} {57003}
  (\bibinfo {year} {2008})}\BibitemShut {NoStop}%
\bibitem [{\citenamefont {Rombouts}\ \emph {et~al.}(1999)\citenamefont
  {Rombouts}, \citenamefont {Heyde},\ and\ \citenamefont
  {Jachowicz}}]{1999PhRvL..82.4155R}%
  \BibitemOpen
  \bibfield  {author} {\bibinfo {author} {\bibfnamefont {S.~M.~A.}\
  \bibnamefont {Rombouts}}, \bibinfo {author} {\bibfnamefont {K.}~\bibnamefont
  {Heyde}}, \ and\ \bibinfo {author} {\bibfnamefont {N.}~\bibnamefont
  {Jachowicz}},\ }\href {\doibase 10.1103/PhysRevLett.82.4155} {\bibfield
  {journal} {\bibinfo  {journal} {Phys. Rev. Lett.}\ }\textbf {\bibinfo
  {volume} {82}},\ \bibinfo {pages} {4155} (\bibinfo {year}
  {1999})}\BibitemShut {NoStop}%
\bibitem [{\citenamefont {Iazzi}\ and\ \citenamefont
  {Troyer}(2015)}]{PhysRevB.91.241118}%
  \BibitemOpen
  \bibfield  {author} {\bibinfo {author} {\bibfnamefont {M.}~\bibnamefont
  {Iazzi}}\ and\ \bibinfo {author} {\bibfnamefont {M.}~\bibnamefont {Troyer}},\
  }\href {\doibase 10.1103/PhysRevB.91.241118} {\bibfield  {journal} {\bibinfo
  {journal} {Phys. Rev. B}\ }\textbf {\bibinfo {volume} {91}},\ \bibinfo
  {pages} {241118} (\bibinfo {year} {2015})}\BibitemShut {NoStop}%
\bibitem [{\citenamefont {Wang}\ \emph
  {et~al.}(2015{\natexlab{b}})\citenamefont {Wang}, \citenamefont {Iazzi},
  \citenamefont {Corboz},\ and\ \citenamefont {Troyer}}]{PhysRevB.91.235151}%
  \BibitemOpen
  \bibfield  {author} {\bibinfo {author} {\bibfnamefont {L.}~\bibnamefont
  {Wang}}, \bibinfo {author} {\bibfnamefont {M.}~\bibnamefont {Iazzi}},
  \bibinfo {author} {\bibfnamefont {P.}~\bibnamefont {Corboz}}, \ and\ \bibinfo
  {author} {\bibfnamefont {M.}~\bibnamefont {Troyer}},\ }\href {\doibase
  10.1103/PhysRevB.91.235151} {\bibfield  {journal} {\bibinfo  {journal} {Phys.
  Rev. B}\ }\textbf {\bibinfo {volume} {91}},\ \bibinfo {pages} {235151}
  (\bibinfo {year} {2015}{\natexlab{b}})}\BibitemShut {NoStop}%
\bibitem [{\citenamefont {Gull}\ \emph {et~al.}(2011)\citenamefont {Gull},
  \citenamefont {Millis}, \citenamefont {Lichtenstein}, \citenamefont
  {Rubtsov}, \citenamefont {Troyer},\ and\ \citenamefont
  {Werner}}]{Gull:2011jd}%
  \BibitemOpen
  \bibfield  {author} {\bibinfo {author} {\bibfnamefont {E.}~\bibnamefont
  {Gull}}, \bibinfo {author} {\bibfnamefont {A.~J.}\ \bibnamefont {Millis}},
  \bibinfo {author} {\bibfnamefont {A.~I.}\ \bibnamefont {Lichtenstein}},
  \bibinfo {author} {\bibfnamefont {A.~N.}\ \bibnamefont {Rubtsov}}, \bibinfo
  {author} {\bibfnamefont {M.}~\bibnamefont {Troyer}}, \ and\ \bibinfo {author}
  {\bibfnamefont {P.}~\bibnamefont {Werner}},\ }\href
  {http://link.aps.org/doi/10.1103/RevModPhys.83.349} {\bibfield  {journal}
  {\bibinfo  {journal} {Rev. Mod. Phys.}\ }\textbf {\bibinfo {volume} {83}},\
  \bibinfo {pages} {349} (\bibinfo {year} {2011})}\BibitemShut {NoStop}%
\bibitem [{\citenamefont {Otsuki}\ \emph {et~al.}(2007)\citenamefont {Otsuki},
  \citenamefont {Kusunose}, \citenamefont {Werner},\ and\ \citenamefont
  {Kuramoto}}]{Otsuki:2007ff}%
  \BibitemOpen
  \bibfield  {author} {\bibinfo {author} {\bibfnamefont {J.}~\bibnamefont
  {Otsuki}}, \bibinfo {author} {\bibfnamefont {H.}~\bibnamefont {Kusunose}},
  \bibinfo {author} {\bibfnamefont {P.}~\bibnamefont {Werner}}, \ and\ \bibinfo
  {author} {\bibfnamefont {Y.}~\bibnamefont {Kuramoto}},\ }\href
  {http://journals.jps.jp/doi/abs/10.1143/JPSJ.76.114707} {\bibfield  {journal}
  {\bibinfo  {journal} {J. Phys. Soc. Jpn.}\ }\textbf {\bibinfo {volume}
  {76}},\ \bibinfo {pages} {114707} (\bibinfo {year} {2007})}\BibitemShut
  {NoStop}%
\bibitem [{\citenamefont {Coqblin}\ and\ \citenamefont
  {Schrieffer}(1969)}]{Coqblin:1969wu}%
  \BibitemOpen
  \bibfield  {author} {\bibinfo {author} {\bibfnamefont {B.}~\bibnamefont
  {Coqblin}}\ and\ \bibinfo {author} {\bibfnamefont {J.~R.}\ \bibnamefont
  {Schrieffer}},\ }\href@noop {} {\bibfield  {journal} {\bibinfo  {journal}
  {Phys. Rev.}\ }\textbf {\bibinfo {volume} {185}},\ \bibinfo {pages} {847}
  (\bibinfo {year} {1969})}\BibitemShut {NoStop}%
\bibitem [{\citenamefont {Yang}\ and\ \citenamefont
  {Lee}(1952)}]{PhysRev.87.404}%
  \BibitemOpen
  \bibfield  {author} {\bibinfo {author} {\bibfnamefont {C.~N.}\ \bibnamefont
  {Yang}}\ and\ \bibinfo {author} {\bibfnamefont {T.~D.}\ \bibnamefont {Lee}},\
  }\href {\doibase 10.1103/PhysRev.87.404} {\bibfield  {journal} {\bibinfo
  {journal} {Phys. Rev.}\ }\textbf {\bibinfo {volume} {87}},\ \bibinfo {pages}
  {404} (\bibinfo {year} {1952})}\BibitemShut {NoStop}%
\bibitem [{\citenamefont {Lee}\ and\ \citenamefont
  {Yang}(1952)}]{PhysRev.87.410}%
  \BibitemOpen
  \bibfield  {author} {\bibinfo {author} {\bibfnamefont {T.~D.}\ \bibnamefont
  {Lee}}\ and\ \bibinfo {author} {\bibfnamefont {C.~N.}\ \bibnamefont {Yang}},\
  }\href {\doibase 10.1103/PhysRev.87.410} {\bibfield  {journal} {\bibinfo
  {journal} {Phys. Rev.}\ }\textbf {\bibinfo {volume} {87}},\ \bibinfo {pages}
  {410} (\bibinfo {year} {1952})}\BibitemShut {NoStop}%
\bibitem [{\citenamefont {Anderson}\ and\ \citenamefont
  {Yuval}(1969)}]{Anderson:1969wl}%
  \BibitemOpen
  \bibfield  {author} {\bibinfo {author} {\bibfnamefont {P.~W.}\ \bibnamefont
  {Anderson}}\ and\ \bibinfo {author} {\bibfnamefont {G.}~\bibnamefont
  {Yuval}},\ }\href
  {http://journals.aps.org/prl/abstract/10.1103/PhysRevLett.23.89} {\bibfield
  {journal} {\bibinfo  {journal} {Phys. Rev. Lett.}\ }\textbf {\bibinfo
  {volume} {23}},\ \bibinfo {pages} {89} (\bibinfo {year} {1969})}\BibitemShut
  {NoStop}%
\bibitem [{\citenamefont {Anderson}\ \emph {et~al.}(1970)\citenamefont
  {Anderson}, \citenamefont {Yuval},\ and\ \citenamefont
  {Hamann}}]{Anderson:1970th}%
  \BibitemOpen
  \bibfield  {author} {\bibinfo {author} {\bibfnamefont {P.~W.}\ \bibnamefont
  {Anderson}}, \bibinfo {author} {\bibfnamefont {G.}~\bibnamefont {Yuval}}, \
  and\ \bibinfo {author} {\bibfnamefont {D.~R.}\ \bibnamefont {Hamann}},\
  }\href {http://journals.aps.org/prb/abstract/10.1103/PhysRevB.1.4464}
  {\bibfield  {journal} {\bibinfo  {journal} {Phys. Rev. B}\ }\textbf {\bibinfo
  {volume} {1}},\ \bibinfo {pages} {4464} (\bibinfo {year} {1970})}\BibitemShut
  {NoStop}%
\bibitem [{\citenamefont {Schotte}(1970)}]{schotte1970tomonaga}%
  \BibitemOpen
  \bibfield  {author} {\bibinfo {author} {\bibfnamefont {K.}~\bibnamefont
  {Schotte}},\ }\href@noop {} {\bibfield  {journal} {\bibinfo  {journal}
  {Zeitschrift f{\"u}r Physik}\ }\textbf {\bibinfo {volume} {230}},\ \bibinfo
  {pages} {99} (\bibinfo {year} {1970})}\BibitemShut {NoStop}%
\bibitem [{\citenamefont {Anderson}\ and\ \citenamefont
  {Yuval}(1973)}]{Anderson:1973hba}%
  \BibitemOpen
  \bibfield  {author} {\bibinfo {author} {\bibfnamefont {P.~W.}\ \bibnamefont
  {Anderson}}\ and\ \bibinfo {author} {\bibfnamefont {G.}~\bibnamefont
  {Yuval}},\ }\href {\doibase 10.1016/B978-0-12-575305-0.50016-8} {\emph
  {\bibinfo {title} {{Asymptotically Exact Methods in the Kondo Problem}}}}\
  (\bibinfo  {publisher} {Academic Press, Inc.},\ \bibinfo {year}
  {1973})\BibitemShut {NoStop}%
\bibitem [{Note1()}]{Note1}%
  \BibitemOpen
  \bibinfo {note} {Ref.~\cite {1971PhRvB...4.2228S} performed one of the first
  historic Carlo simulation of the Kondo model Eq.~(\ref {eq:kondo}) based on
  the Coulomb gas analogy. However the simulation was performed in a canonical
  ensemble with a fixed number of spin-flips, and thus has systematic errors in
  the high temperature region~\cite {Anderson:1973hba}.}\BibitemShut {Stop}%
\bibitem [{\citenamefont {Anderson}(1961)}]{1961PhRv..124...41A}%
  \BibitemOpen
  \bibfield  {author} {\bibinfo {author} {\bibfnamefont {P.~W.}\ \bibnamefont
  {Anderson}},\ }\href
  {http://adsabs.harvard.edu/cgi-bin/nph-data_query?bibcode=1961PhRv..124...41A&link_type=ABSTRACT}
  {\bibfield  {journal} {\bibinfo  {journal} {Phys. Rev.}\ }\textbf {\bibinfo
  {volume} {124}},\ \bibinfo {pages} {41} (\bibinfo {year} {1961})}\BibitemShut
  {NoStop}%
\bibitem [{\citenamefont {Krishna-Murthy}\ \emph {et~al.}(1980)\citenamefont
  {Krishna-Murthy}, \citenamefont {Wilkins},\ and\ \citenamefont
  {Wilson}}]{KrishnaMurthy:1980wb}%
  \BibitemOpen
  \bibfield  {author} {\bibinfo {author} {\bibfnamefont {H.~R.}\ \bibnamefont
  {Krishna-Murthy}}, \bibinfo {author} {\bibfnamefont {J.~W.}\ \bibnamefont
  {Wilkins}}, \ and\ \bibinfo {author} {\bibfnamefont {K.~G.}\ \bibnamefont
  {Wilson}},\ }\href
  {http://journals.aps.org/prb/abstract/10.1103/PhysRevB.21.1003} {\bibfield
  {journal} {\bibinfo  {journal} {Phys. Rev. B}\ }\textbf {\bibinfo {volume}
  {21}},\ \bibinfo {pages} {1003} (\bibinfo {year} {1980})}\BibitemShut
  {NoStop}%
\bibitem [{\citenamefont {Haldane}(1978)}]{1978PhRvL..40..416H}%
  \BibitemOpen
  \bibfield  {author} {\bibinfo {author} {\bibfnamefont {F.}~\bibnamefont
  {Haldane}},\ }\href
  {http://adsabs.harvard.edu/cgi-bin/nph-data_query?bibcode=1978PhRvL..40..416H&link_type=ABSTRACT}
  {\bibfield  {journal} {\bibinfo  {journal} {Phys. Rev. Lett.}\ }\textbf
  {\bibinfo {volume} {40}},\ \bibinfo {pages} {416} (\bibinfo {year}
  {1978})}\BibitemShut {NoStop}%
\bibitem [{\citenamefont {Anderson}(1967)}]{Anderson:1967tr}%
  \BibitemOpen
  \bibfield  {author} {\bibinfo {author} {\bibfnamefont {P.~W.}\ \bibnamefont
  {Anderson}},\ }\href
  {http://journals.aps.org/prl/abstract/10.1103/PhysRevLett.18.1049} {\bibfield
   {journal} {\bibinfo  {journal} {Phys. Rev. Lett.}\ }\textbf {\bibinfo
  {volume} {18}},\ \bibinfo {pages} {1049} (\bibinfo {year}
  {1967})}\BibitemShut {NoStop}%
\bibitem [{\citenamefont {Nishikawa}\ \emph {et~al.}(2012)\citenamefont
  {Nishikawa}, \citenamefont {Crow},\ and\ \citenamefont
  {Hewson}}]{PhysRevB.86.125134}%
  \BibitemOpen
  \bibfield  {author} {\bibinfo {author} {\bibfnamefont {Y.}~\bibnamefont
  {Nishikawa}}, \bibinfo {author} {\bibfnamefont {D.~J.~G.}\ \bibnamefont
  {Crow}}, \ and\ \bibinfo {author} {\bibfnamefont {A.~C.}\ \bibnamefont
  {Hewson}},\ }\href {\doibase 10.1103/PhysRevB.86.125134} {\bibfield
  {journal} {\bibinfo  {journal} {Phys. Rev. B}\ }\textbf {\bibinfo {volume}
  {86}},\ \bibinfo {pages} {125134} (\bibinfo {year} {2012})}\BibitemShut
  {NoStop}%
\bibitem [{\citenamefont {Ruderman}\ and\ \citenamefont
  {Kittel}(1954)}]{PhysRev.96.99}%
  \BibitemOpen
  \bibfield  {author} {\bibinfo {author} {\bibfnamefont {M.~A.}\ \bibnamefont
  {Ruderman}}\ and\ \bibinfo {author} {\bibfnamefont {C.}~\bibnamefont
  {Kittel}},\ }\href {\doibase 10.1103/PhysRev.96.99} {\bibfield  {journal}
  {\bibinfo  {journal} {Phys. Rev.}\ }\textbf {\bibinfo {volume} {96}},\
  \bibinfo {pages} {99} (\bibinfo {year} {1954})}\BibitemShut {NoStop}%
\bibitem [{\citenamefont {Kasuya}(1956)}]{kasuya1956theory}%
  \BibitemOpen
  \bibfield  {author} {\bibinfo {author} {\bibfnamefont {T.}~\bibnamefont
  {Kasuya}},\ }\href@noop {} {\bibfield  {journal} {\bibinfo  {journal}
  {Progress of theoretical physics}\ }\textbf {\bibinfo {volume} {16}},\
  \bibinfo {pages} {45} (\bibinfo {year} {1956})}\BibitemShut {NoStop}%
\bibitem [{\citenamefont {Yosida}(1957)}]{PhysRev.106.893}%
  \BibitemOpen
  \bibfield  {author} {\bibinfo {author} {\bibfnamefont {K.}~\bibnamefont
  {Yosida}},\ }\href {\doibase 10.1103/PhysRev.106.893} {\bibfield  {journal}
  {\bibinfo  {journal} {Phys. Rev.}\ }\textbf {\bibinfo {volume} {106}},\
  \bibinfo {pages} {893} (\bibinfo {year} {1957})}\BibitemShut {NoStop}%
\bibitem [{\citenamefont {Jones}\ and\ \citenamefont
  {Varma}(1987)}]{PhysRevLett.58.843}%
  \BibitemOpen
  \bibfield  {author} {\bibinfo {author} {\bibfnamefont {B.~A.}\ \bibnamefont
  {Jones}}\ and\ \bibinfo {author} {\bibfnamefont {C.~M.}\ \bibnamefont
  {Varma}},\ }\href {\doibase 10.1103/PhysRevLett.58.843} {\bibfield  {journal}
  {\bibinfo  {journal} {Phys. Rev. Lett.}\ }\textbf {\bibinfo {volume} {58}},\
  \bibinfo {pages} {843} (\bibinfo {year} {1987})}\BibitemShut {NoStop}%
\bibitem [{\citenamefont {Jones}\ \emph {et~al.}(1988)\citenamefont {Jones},
  \citenamefont {Varma},\ and\ \citenamefont {Wilkins}}]{PhysRevLett.61.125}%
  \BibitemOpen
  \bibfield  {author} {\bibinfo {author} {\bibfnamefont {B.~A.}\ \bibnamefont
  {Jones}}, \bibinfo {author} {\bibfnamefont {C.~M.}\ \bibnamefont {Varma}}, \
  and\ \bibinfo {author} {\bibfnamefont {J.~W.}\ \bibnamefont {Wilkins}},\
  }\href {\doibase 10.1103/PhysRevLett.61.125} {\bibfield  {journal} {\bibinfo
  {journal} {Phys. Rev. Lett.}\ }\textbf {\bibinfo {volume} {61}},\ \bibinfo
  {pages} {125} (\bibinfo {year} {1988})}\BibitemShut {NoStop}%
\bibitem [{\citenamefont {Affleck}\ \emph {et~al.}(1995)\citenamefont
  {Affleck}, \citenamefont {Ludwig},\ and\ \citenamefont
  {Jones}}]{PhysRevB.52.9528}%
  \BibitemOpen
  \bibfield  {author} {\bibinfo {author} {\bibfnamefont {I.}~\bibnamefont
  {Affleck}}, \bibinfo {author} {\bibfnamefont {A.~W.~W.}\ \bibnamefont
  {Ludwig}}, \ and\ \bibinfo {author} {\bibfnamefont {B.~A.}\ \bibnamefont
  {Jones}},\ }\href {\doibase 10.1103/PhysRevB.52.9528} {\bibfield  {journal}
  {\bibinfo  {journal} {Phys. Rev. B}\ }\textbf {\bibinfo {volume} {52}},\
  \bibinfo {pages} {9528} (\bibinfo {year} {1995})}\BibitemShut {NoStop}%
\bibitem [{\citenamefont {Mitchell}\ \emph {et~al.}(2012)\citenamefont
  {Mitchell}, \citenamefont {Sela},\ and\ \citenamefont
  {Logan}}]{PhysRevLett.108.086405}%
  \BibitemOpen
  \bibfield  {author} {\bibinfo {author} {\bibfnamefont {A.~K.}\ \bibnamefont
  {Mitchell}}, \bibinfo {author} {\bibfnamefont {E.}~\bibnamefont {Sela}}, \
  and\ \bibinfo {author} {\bibfnamefont {D.~E.}\ \bibnamefont {Logan}},\ }\href
  {\doibase 10.1103/PhysRevLett.108.086405} {\bibfield  {journal} {\bibinfo
  {journal} {Phys. Rev. Lett.}\ }\textbf {\bibinfo {volume} {108}},\ \bibinfo
  {pages} {086405} (\bibinfo {year} {2012})}\BibitemShut {NoStop}%
\bibitem [{\citenamefont {Sun}\ and\ \citenamefont
  {Kotliar}(2005)}]{PhysRevLett.95.016402}%
  \BibitemOpen
  \bibfield  {author} {\bibinfo {author} {\bibfnamefont {P.}~\bibnamefont
  {Sun}}\ and\ \bibinfo {author} {\bibfnamefont {G.}~\bibnamefont {Kotliar}},\
  }\href {\doibase 10.1103/PhysRevLett.95.016402} {\bibfield  {journal}
  {\bibinfo  {journal} {Phys. Rev. Lett.}\ }\textbf {\bibinfo {volume} {95}},\
  \bibinfo {pages} {016402} (\bibinfo {year} {2005})}\BibitemShut {NoStop}%
\bibitem [{\citenamefont {Jones}\ and\ \citenamefont
  {Varma}(1989)}]{PhysRevB.40.324}%
  \BibitemOpen
  \bibfield  {author} {\bibinfo {author} {\bibfnamefont {B.~A.}\ \bibnamefont
  {Jones}}\ and\ \bibinfo {author} {\bibfnamefont {C.~M.}\ \bibnamefont
  {Varma}},\ }\href {\doibase 10.1103/PhysRevB.40.324} {\bibfield  {journal}
  {\bibinfo  {journal} {Phys. Rev. B}\ }\textbf {\bibinfo {volume} {40}},\
  \bibinfo {pages} {324} (\bibinfo {year} {1989})}\BibitemShut {NoStop}%
\bibitem [{\citenamefont {Werner}\ and\ \citenamefont
  {Millis}(2006)}]{Werner:2006iza}%
  \BibitemOpen
  \bibfield  {author} {\bibinfo {author} {\bibfnamefont {P.}~\bibnamefont
  {Werner}}\ and\ \bibinfo {author} {\bibfnamefont {A.~J.}\ \bibnamefont
  {Millis}},\ }\href {\doibase 10.1103/PhysRevB.74.155107} {\bibfield
  {journal} {\bibinfo  {journal} {Phys. Rev. B}\ }\textbf {\bibinfo {volume}
  {74}},\ \bibinfo {pages} {155107} (\bibinfo {year} {2006})}\BibitemShut
  {NoStop}%
\bibitem [{Note2()}]{Note2}%
  \BibitemOpen
  \bibinfo {note} {If there is a tunneling between the two impurities, the
  charge transfer will smear out the quantum phase transition into a
  crossover~\cite {PhysRevLett.97.166802}. Even in this case we still identify
  a peak in the fidelity susceptibility (not shown) that indicates the change
  of the system's state in the crossover region.}\BibitemShut {Stop}%
\bibitem [{\citenamefont {Zar\'and}\ \emph {et~al.}(2006)\citenamefont
  {Zar\'and}, \citenamefont {Chung}, \citenamefont {Simon},\ and\ \citenamefont
  {Vojta}}]{PhysRevLett.97.166802}%
  \BibitemOpen
  \bibfield  {author} {\bibinfo {author} {\bibfnamefont {G.}~\bibnamefont
  {Zar\'and}}, \bibinfo {author} {\bibfnamefont {C.-H.}\ \bibnamefont {Chung}},
  \bibinfo {author} {\bibfnamefont {P.}~\bibnamefont {Simon}}, \ and\ \bibinfo
  {author} {\bibfnamefont {M.}~\bibnamefont {Vojta}},\ }\href {\doibase
  10.1103/PhysRevLett.97.166802} {\bibfield  {journal} {\bibinfo  {journal}
  {Phys. Rev. Lett.}\ }\textbf {\bibinfo {volume} {97}},\ \bibinfo {pages}
  {166802} (\bibinfo {year} {2006})}\BibitemShut {NoStop}%
\bibitem [{\citenamefont {Jeong}\ \emph {et~al.}(2001)\citenamefont {Jeong},
  \citenamefont {Chang},\ and\ \citenamefont {Melloch}}]{2001Sci...293.2221J}%
  \BibitemOpen
  \bibfield  {author} {\bibinfo {author} {\bibfnamefont {H.}~\bibnamefont
  {Jeong}}, \bibinfo {author} {\bibfnamefont {A.~M.}\ \bibnamefont {Chang}}, \
  and\ \bibinfo {author} {\bibfnamefont {M.~R.}\ \bibnamefont {Melloch}},\
  }\href
  {http://adsabs.harvard.edu/cgi-bin/nph-data_query?bibcode=2001Sci...293.2221J&link_type=ABSTRACT}
  {\bibfield  {journal} {\bibinfo  {journal} {Science}\ }\textbf {\bibinfo
  {volume} {293}},\ \bibinfo {pages} {2221} (\bibinfo {year}
  {2001})}\BibitemShut {NoStop}%
\bibitem [{\citenamefont {Liu}\ \emph {et~al.}(2010)\citenamefont {Liu},
  \citenamefont {Chandrasekharan},\ and\ \citenamefont
  {Baranger}}]{Liu:2010fz}%
  \BibitemOpen
  \bibfield  {author} {\bibinfo {author} {\bibfnamefont {D.~E.}\ \bibnamefont
  {Liu}}, \bibinfo {author} {\bibfnamefont {S.}~\bibnamefont
  {Chandrasekharan}}, \ and\ \bibinfo {author} {\bibfnamefont {H.~U.}\
  \bibnamefont {Baranger}},\ }\href
  {http://link.aps.org/doi/10.1103/PhysRevLett.105.256801} {\bibfield
  {journal} {\bibinfo  {journal} {Phys. Rev. Lett.}\ }\textbf {\bibinfo
  {volume} {105}},\ \bibinfo {pages} {256801} (\bibinfo {year}
  {2010})}\BibitemShut {NoStop}%
\bibitem [{\citenamefont {Bork}\ \emph {et~al.}(2011)\citenamefont {Bork},
  \citenamefont {Zhang}, \citenamefont {Diekh{\"o}ner}, \citenamefont {Borda},
  \citenamefont {Simon}, \citenamefont {Kroha}, \citenamefont {Wahl},\ and\
  \citenamefont {Kern}}]{Bork:2011bnc}%
  \BibitemOpen
  \bibfield  {author} {\bibinfo {author} {\bibfnamefont {J.}~\bibnamefont
  {Bork}}, \bibinfo {author} {\bibfnamefont {Y.-h.}\ \bibnamefont {Zhang}},
  \bibinfo {author} {\bibfnamefont {L.}~\bibnamefont {Diekh{\"o}ner}}, \bibinfo
  {author} {\bibfnamefont {L.}~\bibnamefont {Borda}}, \bibinfo {author}
  {\bibfnamefont {P.}~\bibnamefont {Simon}}, \bibinfo {author} {\bibfnamefont
  {J.}~\bibnamefont {Kroha}}, \bibinfo {author} {\bibfnamefont
  {P.}~\bibnamefont {Wahl}}, \ and\ \bibinfo {author} {\bibfnamefont
  {K.}~\bibnamefont {Kern}},\ }\href {http://dx.doi.org/10.1038/nphys2076}
  {\bibfield  {journal} {\bibinfo  {journal} {Nature Physics}\ }\textbf
  {\bibinfo {volume} {7}},\ \bibinfo {pages} {901} (\bibinfo {year}
  {2011})}\BibitemShut {NoStop}%
\bibitem [{\citenamefont {Maurand}\ \emph {et~al.}(2012)\citenamefont
  {Maurand}, \citenamefont {Meng}, \citenamefont {Bonet}, \citenamefont
  {Florens}, \citenamefont {Marty},\ and\ \citenamefont
  {Wernsdorfer}}]{Maurand:2012cb}%
  \BibitemOpen
  \bibfield  {author} {\bibinfo {author} {\bibfnamefont {R.}~\bibnamefont
  {Maurand}}, \bibinfo {author} {\bibfnamefont {T.}~\bibnamefont {Meng}},
  \bibinfo {author} {\bibfnamefont {E.}~\bibnamefont {Bonet}}, \bibinfo
  {author} {\bibfnamefont {S.}~\bibnamefont {Florens}}, \bibinfo {author}
  {\bibfnamefont {L.}~\bibnamefont {Marty}}, \ and\ \bibinfo {author}
  {\bibfnamefont {W.}~\bibnamefont {Wernsdorfer}},\ }\href
  {http://link.aps.org/doi/10.1103/PhysRevX.2.011009} {\bibfield  {journal}
  {\bibinfo  {journal} {Phys. Rev. X}\ }\textbf {\bibinfo {volume} {2}},\
  \bibinfo {pages} {011009} (\bibinfo {year} {2012})}\BibitemShut {NoStop}%
\bibitem [{\citenamefont {Mebrahtu}\ \emph {et~al.}(2012)\citenamefont
  {Mebrahtu}, \citenamefont {Borzenets}, \citenamefont {Liu}, \citenamefont
  {Zheng}, \citenamefont {Bomze}, \citenamefont {Smirnov}, \citenamefont
  {Baranger},\ and\ \citenamefont {Finkelstein}}]{Mebrahtu:2012hg}%
  \BibitemOpen
  \bibfield  {author} {\bibinfo {author} {\bibfnamefont {H.~T.}\ \bibnamefont
  {Mebrahtu}}, \bibinfo {author} {\bibfnamefont {I.~V.}\ \bibnamefont
  {Borzenets}}, \bibinfo {author} {\bibfnamefont {D.~E.}\ \bibnamefont {Liu}},
  \bibinfo {author} {\bibfnamefont {H.}~\bibnamefont {Zheng}}, \bibinfo
  {author} {\bibfnamefont {Y.~V.}\ \bibnamefont {Bomze}}, \bibinfo {author}
  {\bibfnamefont {A.~I.}\ \bibnamefont {Smirnov}}, \bibinfo {author}
  {\bibfnamefont {H.~U.}\ \bibnamefont {Baranger}}, \ and\ \bibinfo {author}
  {\bibfnamefont {G.}~\bibnamefont {Finkelstein}},\ }\href
  {http://www.nature.com/doifinder/10.1038/nature11265} {\bibfield  {journal}
  {\bibinfo  {journal} {Nature}\ }\textbf {\bibinfo {volume} {488}},\ \bibinfo
  {pages} {61} (\bibinfo {year} {2012})}\BibitemShut {NoStop}%
\bibitem [{\citenamefont {Franke}\ \emph {et~al.}(2011)\citenamefont {Franke},
  \citenamefont {Schulze},\ and\ \citenamefont {Pascual}}]{Boris:2011cu}%
  \BibitemOpen
  \bibfield  {author} {\bibinfo {author} {\bibfnamefont {K.~J.}\ \bibnamefont
  {Franke}}, \bibinfo {author} {\bibfnamefont {G.}~\bibnamefont {Schulze}}, \
  and\ \bibinfo {author} {\bibfnamefont {J.~I.}\ \bibnamefont {Pascual}},\
  }\href {http://www.sciencemag.org/content/332/6032/940.short} {\bibfield
  {journal} {\bibinfo  {journal} {Science}\ }\textbf {\bibinfo {volume}
  {332}},\ \bibinfo {pages} {940} (\bibinfo {year} {2011})}\BibitemShut
  {NoStop}%
\bibitem [{\citenamefont {Chen}\ \emph {et~al.}(2011)\citenamefont {Chen},
  \citenamefont {Li}, \citenamefont {Cullen}, \citenamefont {Williams},\ and\
  \citenamefont {Fuhrer}}]{Chen:2011jm}%
  \BibitemOpen
  \bibfield  {author} {\bibinfo {author} {\bibfnamefont {J.-H.}\ \bibnamefont
  {Chen}}, \bibinfo {author} {\bibfnamefont {L.}~\bibnamefont {Li}}, \bibinfo
  {author} {\bibfnamefont {W.~G.}\ \bibnamefont {Cullen}}, \bibinfo {author}
  {\bibfnamefont {E.~D.}\ \bibnamefont {Williams}}, \ and\ \bibinfo {author}
  {\bibfnamefont {M.~S.}\ \bibnamefont {Fuhrer}},\ }\href {\doibase
  10.1038/nphys1962} {\bibfield  {journal} {\bibinfo  {journal} {Nature
  Physics}\ }\textbf {\bibinfo {volume} {7}},\ \bibinfo {pages} {535} (\bibinfo
  {year} {2011})}\BibitemShut {NoStop}%
\bibitem [{\citenamefont {Pillet}\ \emph {et~al.}(2013)\citenamefont {Pillet},
  \citenamefont {Joyez}, \citenamefont {\ifmmode~\check{Z}\else
  \v{Z}\fi{}itko},\ and\ \citenamefont {Goffman}}]{PhysRevB.88.045101}%
  \BibitemOpen
  \bibfield  {author} {\bibinfo {author} {\bibfnamefont {J.-D.}\ \bibnamefont
  {Pillet}}, \bibinfo {author} {\bibfnamefont {P.}~\bibnamefont {Joyez}},
  \bibinfo {author} {\bibfnamefont {R.}~\bibnamefont {\ifmmode~\check{Z}\else
  \v{Z}\fi{}itko}}, \ and\ \bibinfo {author} {\bibfnamefont {M.~F.}\
  \bibnamefont {Goffman}},\ }\href {\doibase 10.1103/PhysRevB.88.045101}
  {\bibfield  {journal} {\bibinfo  {journal} {Phys. Rev. B}\ }\textbf {\bibinfo
  {volume} {88}},\ \bibinfo {pages} {045101} (\bibinfo {year}
  {2013})}\BibitemShut {NoStop}%
\bibitem [{\citenamefont {Potok}\ \emph {et~al.}(2007)\citenamefont {Potok},
  \citenamefont {Rau}, \citenamefont {Shtrikman}, \citenamefont {Oreg},\ and\
  \citenamefont {Goldhaber-Gordon}}]{Potok:2007bfa}%
  \BibitemOpen
  \bibfield  {author} {\bibinfo {author} {\bibfnamefont {R.~M.}\ \bibnamefont
  {Potok}}, \bibinfo {author} {\bibfnamefont {I.~G.}\ \bibnamefont {Rau}},
  \bibinfo {author} {\bibfnamefont {H.}~\bibnamefont {Shtrikman}}, \bibinfo
  {author} {\bibfnamefont {Y.}~\bibnamefont {Oreg}}, \ and\ \bibinfo {author}
  {\bibfnamefont {D.}~\bibnamefont {Goldhaber-Gordon}},\ }\href
  {http://www.nature.com/doifinder/10.1038/nature05556} {\bibfield  {journal}
  {\bibinfo  {journal} {Nature}\ }\textbf {\bibinfo {volume} {446}},\ \bibinfo
  {pages} {167} (\bibinfo {year} {2007})}\BibitemShut {NoStop}%
\bibitem [{\citenamefont {Roch}\ \emph {et~al.}(2008)\citenamefont {Roch},
  \citenamefont {Florens}, \citenamefont {Bouchiat}, \citenamefont
  {Wernsdorfer},\ and\ \citenamefont {Balestro}}]{Roch:2008is}%
  \BibitemOpen
  \bibfield  {author} {\bibinfo {author} {\bibfnamefont {N.}~\bibnamefont
  {Roch}}, \bibinfo {author} {\bibfnamefont {S.}~\bibnamefont {Florens}},
  \bibinfo {author} {\bibfnamefont {V.}~\bibnamefont {Bouchiat}}, \bibinfo
  {author} {\bibfnamefont {W.}~\bibnamefont {Wernsdorfer}}, \ and\ \bibinfo
  {author} {\bibfnamefont {F.}~\bibnamefont {Balestro}},\ }\href
  {http://www.nature.com/doifinder/10.1038/nature06930} {\bibfield  {journal}
  {\bibinfo  {journal} {Nature}\ }\textbf {\bibinfo {volume} {453}},\ \bibinfo
  {pages} {633} (\bibinfo {year} {2008})}\BibitemShut {NoStop}%
\bibitem [{\citenamefont {Wang}\ and\ \citenamefont
  {Millis}(2010)}]{Wang:2010dv}%
  \BibitemOpen
  \bibfield  {author} {\bibinfo {author} {\bibfnamefont {X.}~\bibnamefont
  {Wang}}\ and\ \bibinfo {author} {\bibfnamefont {A.~J.}\ \bibnamefont
  {Millis}},\ }\href {\doibase 10.1103/PhysRevB.81.045106} {\bibfield
  {journal} {\bibinfo  {journal} {Phys. Rev. B}\ }\textbf {\bibinfo {volume}
  {81}},\ \bibinfo {pages} {045106} (\bibinfo {year} {2010})}\BibitemShut
  {NoStop}%
\bibitem [{\citenamefont {Goldstein}\ \emph {et~al.}(2010)\citenamefont
  {Goldstein}, \citenamefont {Berkovits},\ and\ \citenamefont
  {Gefen}}]{Goldstein:2010bo}%
  \BibitemOpen
  \bibfield  {author} {\bibinfo {author} {\bibfnamefont {M.}~\bibnamefont
  {Goldstein}}, \bibinfo {author} {\bibfnamefont {R.}~\bibnamefont
  {Berkovits}}, \ and\ \bibinfo {author} {\bibfnamefont {Y.}~\bibnamefont
  {Gefen}},\ }\href {\doibase 10.1103/PhysRevLett.104.226805} {\bibfield
  {journal} {\bibinfo  {journal} {Phys. Rev. Lett.}\ }\textbf {\bibinfo
  {volume} {104}},\ \bibinfo {pages} {226805} (\bibinfo {year}
  {2010})}\BibitemShut {NoStop}%
\bibitem [{\citenamefont {Rau}\ \emph {et~al.}(2010)\citenamefont {Rau},
  \citenamefont {Amasha}, \citenamefont {Oreg},\ and\ \citenamefont
  {Goldhaber-Gordon}}]{Rau:2013vf}%
  \BibitemOpen
  \bibfield  {author} {\bibinfo {author} {\bibfnamefont {I.}~\bibnamefont
  {Rau}}, \bibinfo {author} {\bibfnamefont {S.}~\bibnamefont {Amasha}},
  \bibinfo {author} {\bibfnamefont {Y.}~\bibnamefont {Oreg}}, \ and\ \bibinfo
  {author} {\bibfnamefont {D.}~\bibnamefont {Goldhaber-Gordon}},\ }in\
  \href@noop {} {\emph {\bibinfo {booktitle} {Understanding Quantum Phase
  Transitions}}}\ (\bibinfo  {publisher} {CRC Press},\ \bibinfo {year} {2010})\
  p.\ \bibinfo {pages} {341}\BibitemShut {NoStop}%
\bibitem [{\citenamefont {Hafermann}\ \emph {et~al.}(2013)\citenamefont
  {Hafermann}, \citenamefont {Werner},\ and\ \citenamefont
  {Gull}}]{Hafermann:2013ilc}%
  \BibitemOpen
  \bibfield  {author} {\bibinfo {author} {\bibfnamefont {H.}~\bibnamefont
  {Hafermann}}, \bibinfo {author} {\bibfnamefont {P.}~\bibnamefont {Werner}}, \
  and\ \bibinfo {author} {\bibfnamefont {E.}~\bibnamefont {Gull}},\ }\href
  {http://dx.doi.org/10.1016/j.cpc.2012.12.013} {\bibfield  {journal} {\bibinfo
   {journal} {Computer Physics Communications}\ }\textbf {\bibinfo {volume}
  {184}},\ \bibinfo {pages} {1280} (\bibinfo {year} {2013})}\BibitemShut
  {NoStop}%
\bibitem [{\citenamefont {Bauer}\ \emph {et~al.}(2011)\citenamefont {Bauer},
  \citenamefont {Carr}, \citenamefont {Evertz}, \citenamefont {Feiguin},
  \citenamefont {Freire}, \citenamefont {Fuchs}, \citenamefont {Gamper},
  \citenamefont {Gukelberger}, \citenamefont {Gull}, \citenamefont {Guertler},
  \citenamefont {Hehn}, \citenamefont {Igarashi}, \citenamefont {Isakov},
  \citenamefont {Koop}, \citenamefont {Ma}, \citenamefont {Mates},
  \citenamefont {Matsuo}, \citenamefont {Parcollet}, \citenamefont {Pawlowski},
  \citenamefont {Picon}, \citenamefont {Pollet}, \citenamefont {Santos},
  \citenamefont {Scarola}, \citenamefont {Schollwock}, \citenamefont {Silva},
  \citenamefont {Surer}, \citenamefont {Todo}, \citenamefont {Trebst},
  \citenamefont {Troyer}, \citenamefont {Wall}, \citenamefont {Werner},\ and\
  \citenamefont {Wessel}}]{BBauer:2011tz}%
  \BibitemOpen
  \bibfield  {author} {\bibinfo {author} {\bibfnamefont {B.}~\bibnamefont
  {Bauer}}, \bibinfo {author} {\bibfnamefont {L.~D.}\ \bibnamefont {Carr}},
  \bibinfo {author} {\bibfnamefont {H.~G.}\ \bibnamefont {Evertz}}, \bibinfo
  {author} {\bibfnamefont {A.}~\bibnamefont {Feiguin}}, \bibinfo {author}
  {\bibfnamefont {J.}~\bibnamefont {Freire}}, \bibinfo {author} {\bibfnamefont
  {S.}~\bibnamefont {Fuchs}}, \bibinfo {author} {\bibfnamefont
  {L.}~\bibnamefont {Gamper}}, \bibinfo {author} {\bibfnamefont
  {J.}~\bibnamefont {Gukelberger}}, \bibinfo {author} {\bibfnamefont
  {E.}~\bibnamefont {Gull}}, \bibinfo {author} {\bibfnamefont {S.}~\bibnamefont
  {Guertler}}, \bibinfo {author} {\bibfnamefont {A.}~\bibnamefont {Hehn}},
  \bibinfo {author} {\bibfnamefont {R.}~\bibnamefont {Igarashi}}, \bibinfo
  {author} {\bibfnamefont {S.~V.}\ \bibnamefont {Isakov}}, \bibinfo {author}
  {\bibfnamefont {D.}~\bibnamefont {Koop}}, \bibinfo {author} {\bibfnamefont
  {P.~N.}\ \bibnamefont {Ma}}, \bibinfo {author} {\bibfnamefont
  {P.}~\bibnamefont {Mates}}, \bibinfo {author} {\bibfnamefont
  {H.}~\bibnamefont {Matsuo}}, \bibinfo {author} {\bibfnamefont
  {O.}~\bibnamefont {Parcollet}}, \bibinfo {author} {\bibfnamefont
  {G.}~\bibnamefont {Pawlowski}}, \bibinfo {author} {\bibfnamefont {J.~D.}\
  \bibnamefont {Picon}}, \bibinfo {author} {\bibfnamefont {L.}~\bibnamefont
  {Pollet}}, \bibinfo {author} {\bibfnamefont {E.}~\bibnamefont {Santos}},
  \bibinfo {author} {\bibfnamefont {V.~W.}\ \bibnamefont {Scarola}}, \bibinfo
  {author} {\bibfnamefont {U.}~\bibnamefont {Schollwock}}, \bibinfo {author}
  {\bibfnamefont {C.}~\bibnamefont {Silva}}, \bibinfo {author} {\bibfnamefont
  {B.}~\bibnamefont {Surer}}, \bibinfo {author} {\bibfnamefont
  {S.}~\bibnamefont {Todo}}, \bibinfo {author} {\bibfnamefont {S.}~\bibnamefont
  {Trebst}}, \bibinfo {author} {\bibfnamefont {M.}~\bibnamefont {Troyer}},
  \bibinfo {author} {\bibfnamefont {M.~L.}\ \bibnamefont {Wall}}, \bibinfo
  {author} {\bibfnamefont {P.}~\bibnamefont {Werner}}, \ and\ \bibinfo {author}
  {\bibfnamefont {S.}~\bibnamefont {Wessel}},\ }\href
  {http://iopscience.iop.org/1742-5468/2011/05/P05001} {\bibfield  {journal}
  {\bibinfo  {journal} {J. Stat. Mech.: Theor. Exp.}\ }\textbf {\bibinfo
  {volume} {2011}},\ \bibinfo {pages} {P05001} (\bibinfo {year}
  {2011})}\BibitemShut {NoStop}%
\bibitem [{\citenamefont {Schotte}\ and\ \citenamefont
  {Schotte}(1971)}]{1971PhRvB...4.2228S}%
  \BibitemOpen
  \bibfield  {author} {\bibinfo {author} {\bibfnamefont {K.~D.}\ \bibnamefont
  {Schotte}}\ and\ \bibinfo {author} {\bibfnamefont {U.}~\bibnamefont
  {Schotte}},\ }\href
  {http://adsabs.harvard.edu/cgi-bin/nph-data_query?bibcode=1971PhRvB...4.2228S&link_type=ABSTRACT}
  {\bibfield  {journal} {\bibinfo  {journal} {Phys. Rev. B}\ }\textbf {\bibinfo
  {volume} {4}},\ \bibinfo {pages} {2228} (\bibinfo {year} {1971})}\BibitemShut
  {NoStop}%
\end{thebibliography}%


\end{document}